\documentclass[preprint,showpacs,preprintnumbers,amsmath,amssymb]{revtex4}
\usepackage[utf8]{inputenc}
\usepackage[dvips]{graphicx}% Include figure files
\usepackage{rotating}
\usepackage{hyperref}
\usepackage{subfig}
\usepackage[normalem]{ulem}
\usepackage{color}
\definecolor{orange}{rgb}{1,0.5,0}

\definecolor{red}{rgb}{1,0,0}
\usepackage[T1]{fontenc}
\usepackage{amsmath}
\usepackage{slashed}

\newcommand{\tvect}[2]{%
\ensuremath{\Bigl(\negthinspace\begin{smallmatrix}#1\\#2\end{smallmatrix}\Bigr)}}

\begin{document}

\begin{center}
{\large \bf Effects of Majorana Physics on the UHE $\nu_{\tau}$ Flux Traversing the Earth }\\ \vspace{.9cm}
\end{center}
\author{Luc\'{\i}a Duarte}
\email{lduarte@fing.edu.uy}
 \affiliation{Instituto de F\'{\i}sica, Facultad de Ingenier\'{\i}a,
 Universidad de la Rep\'ublica \\ Julio Herrera y Reissig 565,(11300) 
Montevideo, Uruguay.}

\author{Ismael Romero}
\affiliation{Instituto de Investigaciones F\'{\i}sicas de Mar del Plata (IFIMAR)\\
CONICET, UNMDP\\ Departamento de F\'{\i}sica,
Universidad Nacional de Mar del Plata \\
Funes 3350, (7600) Mar del Plata, Argentina}

\author{Gabriel Zapata}
\affiliation{Instituto de Investigaciones F\'{\i}sica de Mar del Plata (IFIMAR)\\
CONICET, UNMDP\\ Departamento de F\'{\i}sica,
Universidad Nacional de Mar del Plata \\
Funes 3350, (7600) Mar del Plata, Argentina}
 
\author{Oscar A. Sampayo}
\email{sampayo@mdp.edu.ar}

 \affiliation{Instituto de Investigaciones F\'{\i}sica de Mar del Plata (IFIMAR)\\ CONICET, UNMDP\\ Departamento de F\'{\i}sica,
Universidad Nacional de Mar del Plata \\
Funes 3350, (7600) Mar del Plata, Argentina}

\begin{abstract}
{\small We study the effects produced by sterile Majorana neutrinos on the $\nu_{\tau}$ flux traversing the Earth, considering the interaction 
between the Majorana neutrinos and the standard matter as modeled by an effective theory. The surviving tau-neutrino flux is calculated using transport equations including Majorana neutrino production and decay. We compare our results with the pure Standard Model interactions, computing the surviving flux for different values of the effective lagrangian couplings, considering the detected flux by IceCube for an operation time of ten years, and Majorana neutrinos with mass $m_N \thicksim m_{\tau}$. }
\end{abstract}

\maketitle

%\narrowtext
\section{\bf Introduction}

The discovery of neutrino flavor oscillations still remains as one of the most compelling evidence for physics beyond the Standard Model ($SM$). While many proposals have been posed to explain the tiny ordinary neutrino masses, the seesaw mechanism stays as one of the most straightforward ideas for solving the neutrino mass problem \cite{Minkowski:1977sc, Mohapatra:1979ia, Yanagida:1980xy, GellMann:1980vs, Schechter:1980gr, Kayser:1989iu}. This mechanism introduces right handed sterile neutrinos that, as they do not have distinct particle and antiparticle degrees of freedom, can have a Majorana mass term leading to the tiny known masses for the standard neutrinos, as long as the Yukawa couplings between the right handed Majorana neutrinos and the standard ones remain small. For Yukawa couplings of order $Y \sim 1$, we need a Majorana mass scale of order $M_{N} \sim 10^{15} GeV$ to account for a light $\nu$ mass compatible with the current neutrino data ($m_{\nu}\sim 0.01 e$V), and this fact leads to the decoupling of the Majorana neutrinos. On the other hand, for smaller Yukawa couplings, of the order $Y\sim 10^{-8}-10^{-6}$, sterile neutrinos with masses around $M_{N}\sim (1-1000) ~GeV$ could exist, but in the simplest Type-I seesaw scenario with sterile Majorana neutrinos, this leads to a negligible left-right neutrino mixing $U_{lN}^2 \sim m_{\nu}/M_N \sim 10^{-14}-10^{-10}$ \cite{deGouvea:2015euy, Deppisch:2015qwa, delAguila:2008ir}. Thus, as suggested in \cite{delAguila:2008ir}, the detection of Majorana neutrinos ($N$) would be a signal of physics beyond the minimal seesaw mechanism, and its interactions could be better described in a model independent approach based on an effective theory, considering a scenario with only one Majorana neutrino $N$ and negligible mixing with the $\nu_{L}$.

On the other hand, in the recent years the observation of ultra high energy (UHE) astrophysical neutrinos in the IceCube telescope \cite{Aartsen:2014gkd}, with a yet unknown specific origin, spectral shape and flavor composition, as well as the non-finding of tau neutrinos in still primary searches performed within these data \cite{Aartsen:2015dlt}, raise the question on tau-neutrino detection in neutrino telescopes.

In addition, anomalies found in short baseline ($SBL$) neutrino oscillation experiments \cite{Aguilar:2001ty, AguilarArevalo:2007it, AguilarArevalo:2008rc, Mention:2011rk} have driven the introduction of light, almost sterile neutrinos, that mix poorly with the known light mass states and could help to accommodate the data introducing a third mass splitting. The IceCube collaboration has searched for these sterile neutrinos \cite{TheIceCube:2016oqi} probing light sterile neutrino $3+1$ models \cite{Nunokawa:2003ep, Kopp:2013vaa}, and recently led to new bounds for the sterile-active muon neutrino mixing.  

As the mixing parameters for the second fermion family with a sterile Majorana neutrino $\nu_{\mu}-\nu_{s}$ are strongly constrained within the framework of $3+1$ scenarios \cite{deGouvea:2015euy, Collin:2016aqd}, and motivated by the lack of tau neutrinos in the UHE cosmic flux in IceCube, in this work we study the possibilities that UHE tau-neutrinos from astrophysical sources may provide a signature for Majorana neutrino production by giving a surviving flux after traversing the Earth which may differ from the standard one. In particular the non-observation of $\nu_{\tau}$ going up signal could be a manifestation of a sterile neutrino modifying the $\nu_{\tau}$ flux.

We study the possibility that the existence of Majorana neutrinos coupled to the tau neutrinos modifies their interactions with nucelons in matter, and thus change the surviving $\nu_{\tau}$ flux after traversing the Earth. We have studied the bounds on the effective Majorana neutrino couplings strenght obtained from different experimental data, and we find this couplings can have appreciable effects on the $\nu_{\tau}$ flux attenuation at high energies. This fact may have an impact on the detection of astrophysical $\nu_{\tau}$ flux. IceCube has recently analyzed high energy neutrino events \cite{Aartsen:2015ivb, Aartsen:2015knd}, and found, although with large uncertainty, consistency with equal fractions of all flavors, but without including a specific tau neutrino identification algorithm. If $\nu_{\tau}$ events are finally found, with the increase of detection time increasing the statistics, then the data can be used to place bounds on the heavy Majorana neutrino effects we are showing in this work.

Many recent papers have studied the IceCube UHE astrophysical events with explanations involving dark matter models with right-handed neutrinos \cite{Fiorentin:2016avj, Dev:2016qbd, Chianese:2016smc, Boucenna:2015tra} and non standard effective interactions \cite{Gonzalez-Garcia:2016gpq}, and also the chances to probe sterile- tau neutrino mixings have been considered \cite{Esmaili:2013cja}. The effects of dark matter and new physics on $\nu_{\tau}$ propagation has been studied by our group in refs. \cite{Reynoso:2012sy, Reynoso:2013wea, Reynoso:2016hjr}, and the Majorana neutrino effective phenomenology regarding the relevant $N$ decay modes and interactions is treated in \cite{Duarte:2015iba, Duarte:2016miz}.     

In Sec.\ref{modelo}, we briefly describe the effective operator approach and the bounds on the effective couplings we take into account. In Sec.\ref{procesos} we discuss the relevant processes and the results obtained for the cross-sections and decay rates. In Sec.\ref{transporte}, we review the passage of high energy tau neutrinos through the Earth using transport equations including the effects of Majorana neutrinos. We solve these equations taking into account the neutral current regeneration and the regeneration by the decay of the Majorana neutrino, for different values of $\zeta$ which, as we will define in the next section, is a combination of the effective couplings and the energy scale associated with the new interactions. This enables us to compare the surviving flux with the one obtained using $SM$ physics only. In particular we go to a Majorana mass region where the couplings are less constrained maximizing the effect on the survival flux. Finally, in Sec.\ref{resultados} we present the results and in Sec.\ref{conclusiones} a short discussion with our conclusions.

\section{\bf Majorana neutrino interaction model \label{modelo}}

\subsection{Effective operators and lagrangian}

In this paper we study the effects of the possible existence of a heavy sterile Majorana neutrino $N$ in the $\nu_{\tau}$ propagation through the Earth. 
Being the $N$ a $SM$  singlet, its only possible renormalizable interactions with $SM$ fields involve the Yukawa couplings. But as we discussed in the introduction, these couplings must be very small in order to accommodate the observed tiny ordinary $\nu$ masses. In this work we take an alternative approach, considering that the sterile $N$ interacts with the light neutrinos by higher dimension effective operators, and take this interaction to be dominant in comparison with the the mixing through the Yukawa couplings. In this sense we depart from the usual viewpoint in which the sterile neutrinos mixing with the standard neutrinos is assumed to govern the $N$ production and decay mechanisms \cite{Atre:2009rg, delAguila:2007qnc}.   

We parameterize the effects of new physics by a  set of effective operators $\mathcal{O}$ constructed with the standard model and the Majorana neutrino fields and satisfying the $SU(2)_L \otimes U(1)_Y$ gauge symmetry \cite{delAguila:2008ir}.  The effect of these operators is suppressed by inverse powers of the new physics scale $\Lambda$, which is not necessarily related to the Majorana neutrino mass $m_{N}$. The total lagrangian is organized as follows:
\begin{eqnarray}
\mathcal{L}=\mathcal{L}_{SM}+\sum_{n=6}^{\infty}\frac1{\Lambda^{n-4}}\sum_i \alpha_i \mathcal{O}_i^{(n)}
\end{eqnarray}
For the considered operators we follow \cite{delAguila:2008ir} starting with a rather general effective lagrangian density for
the interaction of right handed Majorana neutrinos $N$ with bosons, leptons and quarks. We list the dimension $6$ operators that can be generated at tree level or one-loop level in the unknown fundamental ultraviolet theory, and are baryon-number conserving.
The first subset includes operators with scalar and vector bosons (SVB),
\begin{eqnarray} \label{eq:ope1}
\mathcal{O}_{LN\phi}=(\phi^{\dag}\phi)(\bar L N \tilde{\phi}), \;\; \mathcal{O}_{NN\phi}=i(\phi^{\dag}D_{\mu}\phi)(\bar N
\gamma^{\mu} N), \;\; \mathcal{O}_{Ne\phi}=i(\phi^T \epsilon D_{\mu} \phi)(\bar N \gamma^{\mu} l)
\end{eqnarray}
and a second subset includes the baryon-number conserving 4-fermion contact terms:
\begin{eqnarray} \label{eq:ope2}
\mathcal{O}_{duNe}&=&(\bar d \gamma^{\mu} u)(\bar N \gamma_{\mu} l) , \;\; \mathcal{O}_{fNN}=(\bar f \gamma^{\mu}
f)(\bar N \gamma_{\mu}
N), \;\; \mathcal{O}_{LNLe}=(\bar L N)\epsilon (\bar L l),
\nonumber \\
\mathcal{O}_{LNQd}&=&(\bar L N) \epsilon (\bar Q
d), \;\; \mathcal{O}_{QuNL}=(\bar Q u)(\bar N L) , \;\; \mathcal{O}_{QNLd}=(\bar Q N)\epsilon (\bar L d), 
\nonumber \\
\mathcal{O}_{LN}&=&|\bar N L|^2 , \;\; \mathcal{O}_{QN}=|\bar Q N|^2
\end{eqnarray}
where $l$, $u$, $d$ and $L$, $Q$ denote, the right handed $SU(2)$ singlet and the
left-handed $SU(2)$ doublets, respectively.
The following one-loop level generated operators coefficients are naturally suppressed by a factor $1/16\pi^2$ \cite{delAguila:2008ir, Arzt:1994gp}:
\begin{eqnarray}
\mathcal{O}^{(5)}_{NNB} & = & \bar N \sigma^{\mu\nu} N^c B_{\mu\nu}, \cr \mathcal{O}_{ N B} = (\bar L \sigma^{\mu\nu} N) \tilde
\phi B_{\mu\nu} , && \mathcal{O}_{ N W } = (\bar L \sigma^{\mu\nu} \tau^I N) \tilde \phi W_{\mu\nu}^I , \cr \mathcal{O}_{ D N} =
(\bar L D_\mu N) D^\mu \tilde \phi, && \mathcal{O}_{ \bar D N} = (D_\mu \bar L N) D^\mu \tilde \phi \ . \label{eq:ope3}
\end{eqnarray}

In order to study the effects on the $\nu_{\tau}$ propagation through the Earth due to the existence of Majorana neutrinos $N$,  we consider the dominant processes responsible for the change in the $\nu$, $\tau$ and $N$ fluxes. Besides the $SM$ processes involving ordinary neutrinos, we have new contributions related with the production and scattering of the Majorana neutrinos $N$ interacting with matter nucleons ($\mathcal{N}$):

\begin{eqnarray}\label{reacciones}
\nu \mathcal{N} \rightarrow N X, \;~~~~\;
N \mathcal{N} \rightarrow l X,  \;~~~~\;
N \mathcal{N} \rightarrow \nu X,  \;~~~~\;
N \mathcal{N} \rightarrow N X
\end{eqnarray}
Also, we will take into account the Majorana neutrino decay contribution to the different fluxes. For the low Majorana neutrino mass region, the dominant decay was found to be  $N \rightarrow \gamma \; \nu$  \cite{Duarte:2015iba}. For completeness we include in Fig.\ref{fig:branching} a plot with the $N$ Branching Ratios in the low mass region.

The shown reactions \eqref{reacciones} contribute to different terms in the transport equations to be presented in Sec.\ref{transporte}, where the relative relevance of the different terms for the considered mass region will be discussed. 

In order to obtain the above interactions we derive the effective lagrangian terms involved in the calculations, taking the scalar doublet after spontaneous symmetry breaking as $\phi=\tvect{0}{\frac{v+h}{\sqrt{2}}}$. We have contributions to the effective lagrangian coming from \eqref{eq:ope1}, related to the spontaneous symmetry breaking process:
\begin{eqnarray}\label{leff_svb}
 \mathcal{L}^{tree}_{SVB} &=& \frac{1}{\Lambda^2}\left\{
\alpha_Z (\bar N_R \gamma^{\mu} N_R) \left( \frac{v m_Z}{2} Z_{\mu} \right)  \right.
 - \left. \alpha^{(i)}_W (\bar N_R \gamma^{\mu} l_{R,i})\left(\frac{v m_{W}}{\sqrt{2}}W^{+}_{\mu} \right) + \cdots + h.c. \right\}, 
\end{eqnarray}
and the four-fermion interactions involving quarks and leptons from (\ref{eq:ope2})
\begin{eqnarray}\label{leff_4-f}
\mathcal{L}^{tree}_{4-f}&=& \frac{1}{\Lambda^2} \left\{ \alpha^{(i,j)}_{V_0} \bar d_{R,i} \gamma^{\mu} u_{R,i} \bar N_R
\gamma_{\mu} l_{R,j} + \alpha^{(i)}_{V_1} \bar l_{R,i} \gamma^{\mu} l_{R,i} \bar N_R \gamma_{\mu} N_R + \alpha^{(i)}_{V_2} \bar
L_i \gamma^{\mu} L_i \bar N_R \gamma_{\mu} N_R + \right. \nonumber
\\ && \left. \alpha^{(i)}_{V_3} \bar u_{R,i} \gamma^{\mu}
u_{R,i} \bar N_R \gamma_{\mu} N_R + \alpha^{(i)}_{V_4} \bar d_{R,i} \gamma^{\mu} d_{R,i} \bar N_R \gamma_{\mu} N_R +
\alpha^{(i)}_{V_5} \bar Q_i \gamma^{\mu} Q_i \bar N_R \gamma_{\mu} N_R + \right. \nonumber
\\ && \left.
\alpha^{(i,j)}_{S_0}(\bar \nu_{L,i}N_R \bar e_{L,j}l_{R,j}-\bar e_{L,i}N_R \bar \nu_{L,j}l_{R,j}) + \alpha^{(i,j)}_{S_1}(\bar
u_{L,i}u_{R,i}\bar N \nu_{L,j}+\bar d_{L,i}u_{R,i} \bar N e_{L,j})
 + \right. \nonumber
\\ && \left.
\alpha^{(i,j)}_{S_2} (\bar \nu_{L,i}N_R \bar d_{L,j}d_{R,j}-\bar e_{L,i}N_R \bar u_{L,j}d_{R,j}) + \alpha^{(i,j)}_{S_3}(\bar
u_{L,i}N_R \bar e_{L,j}d_{R,j}-\bar d_{L,i}N_R \bar \nu_{L,j}d_{R,j}) + \right. \nonumber
\\ && \left.  \alpha^{(i,j)}_{S_4} (\bar N_R \nu_{L,i}~\bar l_{L,j} N_R~+\bar
N_R e_{L,i} \bar e_{L,j} N_R) + \cdots  + h.c. \right\}
\end{eqnarray}
In Eqs. \eqref{leff_svb} and \eqref{leff_4-f} a sum over the family index $i,j$ is understood, and the constants
$\alpha^{(i,j)}_{\mathcal O}$ are associated to specific operators:
\begin{eqnarray}
\alpha_Z&=&\alpha_{NN\phi},\; \alpha^{(i)}_{\phi}=\alpha^{(i)}_{LN\phi},\; \alpha^{(i)}_W=\alpha^{(i)}_{Ne\phi},\;
\alpha^{(i,j)}_{V_0}=\alpha^{(i,j)}_{duNe},\;\;
\alpha^{(i)}_{V_1}=\alpha^{(i)}_{eNN},\;\nonumber \\
\alpha^{(i)}_{V_2}&=&\alpha^{(i)}_{LNN},\;\alpha^{(i)}_{V_3}=\alpha^{(i)}_{uNN},\;
\alpha^{(i)}_{V_4}=\alpha^{(i)}_{dNN},\;\alpha^{(i)}_{V_5}=\alpha^{(i)}_{QNN},\;
\alpha^{(i,j)}_{S_0}=\alpha^{(i,j)}_{LNe},\;\nonumber \\
\alpha^{(i,j)}_{S_1}&=&\alpha^{(i,j)}_{QuNL},\; \alpha^{(i,j)}_{S_2}=\alpha^{(i,j)}_{LNQd},\;\;
\alpha^{(i,j)}_{S_3}=\alpha^{(i,j)}_{QNLd},\; \alpha^{(i)}_{S_4}=\alpha^{(i)}_{LN}.
\end{eqnarray}
In this work we allow for family mixing in the interaction involving two or more different $SM$ leptons.

The one-loop generated operators are suppressed by the $1/(16 \pi^2)$ factor but, as we show in \cite{Duarte:2015iba}, these play a major role in the
$N$-decay. In particular for the low $m_N$ range studied here, the dominant channel $N \rightarrow \nu \gamma$ is produced by terms coming from the operators in \eqref{eq:ope3}
\begin{eqnarray}\label{leff_1loop}
\mathcal{L}_{eff}^{1-loop}&=&\frac{\alpha_{L_1}^{(i)}}{\Lambda^2} \left(-i\sqrt{2} v c_W P^{(A)}_{\mu} ~\bar \nu_{L,i} \sigma^{\mu\nu} N_R~ A_{\nu} 
+i \sqrt{2} v s_W P^{(Z)}_{\mu} ~\bar \nu_{L,i} \sigma^{\mu\nu} N_R~ Z_{\nu}+  \right)  
\nonumber 
\\ &-& \frac{\alpha_{L_2}^{(i)}}{\Lambda^2} \left(\frac{m_Z}{\sqrt{2}}P^{(N)}_{\mu} ~\bar \nu_{L,i} N_R~ Z^{\mu}+ 
+ m_W P^{(N)}_{\mu} ~\bar l_{L,i} N_R~ W^{-\mu} \right) 
\nonumber 
\\ &-& \frac{\alpha_{L_3}^{(i)}}{\Lambda^2}\left(i\sqrt{2} v  c_W P^{(Z)}_{\mu} ~\bar \nu_{L,i} \sigma^{\mu\nu}N_R~ Z_{\nu} 
+ i\sqrt{2} v s_W P^{(A)}_{\mu} ~\bar \nu_{L,i} \sigma^{\mu\nu}N_R~ A_{\nu} \right.
\nonumber 
\\ &+& \left. i 2\sqrt{2} m_W ~\bar \nu_{L,i} \sigma^{\mu\nu} N_R~ W^+_{\mu}W^-_{\nu} + i \sqrt{2} v P^{(W)}_{\mu} ~\bar l_{L,i} \sigma^{\mu\nu} N_R~ W^-_{\nu} \right.
\nonumber
\\ &+& \left. i 4 m_W c_W ~\bar l_{L,i} \sigma^{\mu\nu} N_R~ W^-_{\mu} Z_{\nu}+ i 4 m_W s_W ~\bar l_{L,i} \sigma^{\mu\nu} N_R~ W^-_{\mu} A_{\nu} 
\right)
\nonumber
\\ &-&  \frac{\alpha_{L_4}^{(i)}}{\Lambda^2} \left( \frac{m_Z}{\sqrt{2}} P^{(\bar\nu)}_{\mu}~\bar \nu_{L,i} N_R~ Z_{\mu}- 
\frac{\sqrt{2} m^2_W}{v} ~\bar \nu_{L,i} N_R~  W^{-\mu}W^+_{\mu} 
-\frac{m^2_z}{\sqrt{2} v} ~\bar \nu_{L,i} N_R~ Z_{\mu}Z^{\mu} \right.
\nonumber
\\ &+& \left.  m_W P^{(\bar l)}_{\mu} W^{-\mu} ~\bar l_{L,i} N_R + 
 e m_W  ~\bar l_{L,i} N_R W^{-\mu}A_{\mu} + e m_Z s_W ~\bar l_{L,i} N_R W^{-\mu}Z_{\mu} 
\right) + h.c.
\end{eqnarray}
where $P^{(a)}$ is the 4-moment of the incoming $a$-particle and a sum over the family index $i$ is understood again.
The constants $\alpha^{(i)}_{L_j}$ with $j=1,3$ are associated to the specific operators:
\begin{eqnarray}
\alpha^{(i)}_{L_1}=\alpha^{(i)}_{NB},\;\; \alpha^{(i)}_{L_2}=\alpha^{(i)}_{DN},\;\; \alpha^{(i)}_{L_3}=\alpha^{(i)}_{NW},\;\; 
\alpha^{(i)}_{L_4}=\alpha^{(i)}_{\bar DN}.
\end{eqnarray} 
The complete lagrangian for the effective model is presented in an appendix in our recent work \cite{Duarte:2016miz}.

In order to maintain the discussion as simple as possible we will consider the contributions of the different operators by sets corresponding to 
$\mathcal{O}_{SVB}$, $\mathcal{O}_{4-f}$, $\mathcal{O}_{1-loop}$ with the couplings $\alpha_{SVB}$, $\alpha_{4-f}$, $\alpha_{1-loop}$ respectively.

\subsection{Experimental bounds on the effective couplings}
 
Existent bounds on right-handed sterile Majorana neutrinos are usually imposed on the parameters representing the mixing between them and the light ordinary neutrinos. Recent works \cite{deGouvea:2015euy, Deppisch:2015qwa, Antusch:2015mia} summarize in general phenomenological approaches the
existing experimental bounds for a sterile neutrino coupled to the three fermion families, considering low scale minimal seesaw models, parameterized 
by a single heavy neutrino mass scale $M_{N}$ and light-heavy mixings $U_{lN}$, with $l$ indicating the lepton flavor. These mixings are constrained
experimentally, depending of the flavor and the decay channels taken into account, by neutrinoless double beta decay, electroweak precision tests, low energy observables as rare lepton number violating (LNV) decays of mesons, peak searches in meson decays and beam dump experiments, as well as direct collider searches involving Z decays.
In the effective lagrangian framework we are studying, the heavy Majorana neutrino couples to the three fermion family flavors with couplings dependent on the new ultraviolet physics scale $\Lambda$ and the constants $\alpha^{(i)}_{\mathcal O}$ associated to the different operators. 
The current experimental bounds on the $U_{lN}$ mixings can be re-interpreted in terms of the effective couplings considering a particular combination of the couplings and the new physics scale that we call $\zeta_{\mathcal{O}}$:
\begin{eqnarray}\label{U2}
\zeta_{\mathcal{O}}=\left(\frac{\alpha_{\mathcal{O}}v^2}{2 \Lambda^2}\right)^2
\end{eqnarray}
where $v=250$ G$e$V represents the Higgs field vacuum expectation value.
Previous analysis \cite{delAguila:2006bda, delAguila:2008iz} refer in general to similar heavy neutrino-standard boson interaction structures that modify the weak currents and lead to variations in the weak bosons decay rates, and $W$ and $Z$ mediated processes involved in the existing experimental tests:
\begin{eqnarray}
\label{lw}
 \mathcal L_W = -\frac{g}{\sqrt{2}}  \overline l
\gamma^{\mu} U_{lN} P_L N W_{\mu} + h.c.
\end{eqnarray}
\begin{eqnarray}
\label{lz}
 \mathcal L_Z = -\frac{g}{2 c_{W}} \overline \nu_{L}
\gamma^{\mu} U_{lN} P_L N Z_{\mu} + h.c.
\end{eqnarray}

As we mentioned above, we consider three sets of operators called $\mathcal{O}_{SVB}$, $\mathcal{O}_{4-f}$, $\mathcal{O}_{1-loop}$ and the existent bounds on their values. The $\mathcal{O}_{SVB}$ operators in \eqref{eq:ope1} lead to a term in the effective lagrangian \eqref{leff_svb} that can be compared to the interaction in \eqref{lw}. The relation between the coupling $\alpha^{(i)}_{W}$ and the mixing $U_{lN}$ was derived in \cite{delAguila:2008ir}: $ U_{lN} \simeq \left(\frac{\alpha^{(i)}_{W}v^2}{2\Lambda^2}\right)$. In our current notation this would be: $ U^{2}_{lN} \simeq \zeta_{SVB}$. 

\begin{table}\small 
\begin{center}
\begin{tabular}{ | c | c | c | c |}
\hline
Process / Coupling 			& $\zeta_{SVB}$ 	& $\zeta_{4-f}$ 	& $\zeta_{1-loop}$ \\ \hline
$Z\rightarrow N N$ \dag 		&$< 7.56 \times 10^{-4}$& - 			& - \\ \hline
$e^{+} e^{-}\rightarrow \nu N$  \dag	& - 			& $<2.85\times 10^{-1}$ 	& - \\ \hline
$e^{+} e^{-}\rightarrow N N$ \dag	& - 			& $<2.63\times 10^{-1}$ 	& - \\ \hline
$Z \rightarrow \nu N$ \dag 		& - 			& - 			& $<6.75\times 10^{-4}$ \\ \hline
%$\tau$ universality tests * 		& $<974$ \ddag		& $>-650$ \ddag		& - \\ \hline
\end{tabular}
\caption{ Experimental bounds on the effective couplings. \break
{\footnotesize{\dag LEP Ref. \cite{Decamp:1991uy} }}} 
%* Refs. \cite{deGouvea:2015euy,Agashe:2014kda}
%\ddag values for $m_N = 0.99~ m_{\tau}$ in Eq.(\ref{eq:rtau}) 
\label{tab:bounds}
\end{center}
\end{table}

The couplings $\zeta_{SVB}$ can be bounded taking into account LEP and $\tau$ lepton universality tests results.  
We consider the LEP bounds on single $Z\rightarrow \nu N$ and pair $Z\rightarrow N~N$ sterile neutrino production searches \cite{Decamp:1991uy}. Conservative limits for any $m_N$ mass \cite{Decamp:1991uy} are
\begin{eqnarray}\label{eq:zNN}
Br(Z\rightarrow NN)Br^{2}(N\rightarrow \nu (\bar \nu) \gamma)<5\times 10^{-5}
\end{eqnarray}
\begin{eqnarray}\label{eq:znuN}
Br(Z\rightarrow \nu N)Br(N\rightarrow \nu (\bar \nu) \gamma)<2.7\times 10^{-5}
\end{eqnarray}
This result is model-independent and holds for the production of a single and a pair of heavy neutral objects decaying into a photon and a light invisible particle.

For the decay $Z\rightarrow N~N$, we have a direct contribution from the tree level operator $\mathcal{O}_{NN\phi}$, giving 
\begin{eqnarray}
 \Gamma(Z\rightarrow N N)= \frac1{24\pi}\left(\frac{\alpha_Z v^2}{2 \Lambda^2} \right)^2 \frac{m_Z^3}{v^2}
\end{eqnarray}

For the low $m_{N}$ values considered in this work, we can take $Br(N\rightarrow \nu (\bar \nu) \gamma)\simeq 1$ and then the corresponding bound is 
\begin{eqnarray} \label{alfaz-bound}
\zeta_{Z}< 7.56 \times 10^{-4}. 
\end{eqnarray}
The process $Z \rightarrow \nu N$ has no contributions from the operators $\mathcal{O}_{SVB}$.

Another observable that can put restrictive bounds on the $\mathcal{O}_{SVB}$ is the universality test from the $\tau$-decay, in the mass range $m_{\mu} \leq m_N \leq m_{\tau}$.
Following \cite{deGouvea:2015euy} we define the quotient $R_{\tau}$ in the effective model as
\begin{eqnarray}\label{eq:rtau}
R_{\tau}^{eff}&=&\frac{\Gamma(\tau \rightarrow \nu_{\tau}(N)+e\bar\nu)}{\Gamma(\mu \rightarrow \nu_{\mu}+e\bar\nu)}=
\frac{\Gamma(\tau \rightarrow \nu_{\tau}+e\bar\nu)+\Gamma(\tau \rightarrow N+e\bar\nu)}
{\Gamma(\mu \rightarrow \nu_{\mu}+e\bar\nu)}
\nonumber \\
&& = \left(\frac{m_{\tau}}{m_{\mu}} \right)^5 \left(\frac{g(y_{\tau})}{g(y_{\mu})}+\frac{h(y_N,\zeta_{S0}, \zeta_{W})}{8g(y_{\mu})}\right)
\end{eqnarray}
The function $g(x)=1-8x+8x^3-x^4-12x^2 \ln(x)$ is the SM result for the $\tau \rightarrow \nu_{\tau}+ e \nu$ and $\mu \rightarrow \nu_{\mu}+ e \nu$ decays, with $y_{\tau}=(m_e/m_{\tau})^2$ and $y_{\mu}=(m_e/m_{\mu})^2$ respectively. The $\tau \rightarrow N + e \nu$ process receives $\mathcal{O}_{4-f}$ and $\mathcal{O}_{SVB}$ contributions, encoded in the function
\begin{eqnarray}
 h(y_N, \zeta_{S0}, \zeta_{W}) & = & (1-y_N^2)(9 \zeta_{S0}(1+y_N^2)+12~ \zeta_{S0}(y_N-1)^2+2 \zeta_{W}(y_N^2-8 y_N+1))
 \nonumber \\
 &&
 +12 ~y_N~ \ln(y_N) (3\zeta_{S0}- 2 \zeta_{W})
\end{eqnarray}
with $y_{N}=(m_N/m_{\tau})^2$. 
The observed value for the quotient is $R_{\tau}^{obs}=(1.349 \pm 0.004)\times 10^{6}$ \cite{deGouvea:2015euy,Agashe:2014kda}. This imposes stringent bounds on both the $SVB$ $\zeta_{W}$ and the 4-fermion $\zeta_{S0}$ couplings, but if we take the mass of the sterile $N$ to be right below $m_{\tau}$, these bounds can be relaxed because the partial decay width is kinematically canceled. 

We study now the case of the effective 4-fermion interactions. Here we have again contributions to the LEP process $e^- e^+ \rightarrow \nu N$ and 
$e^- e^+ \rightarrow N N$ but, in our case without the $Z$ resonance, and then we expect weaker bounds than those imposed on the $\zeta_{SVB}$ couplings. 

We consider first the reaction $e^+ e^- \rightarrow \nu N$ calculated at the $Z$-pole 
\begin{eqnarray}
\sigma_{\nu N}=\frac{12\pi}{m_Z^2}Br(Z\rightarrow e^+e^-)Br(Z\rightarrow \nu N)
\end{eqnarray}
where $Br(Z\rightarrow e^+ e^-)=3.4\times 10^{-2}$ and the upper bound for the Branching ratio for the channel $Z\rightarrow \nu N$ obtained from \eqref{eq:znuN} is $Br(Z\rightarrow \nu N)\leq 2.7 \times 10^{-5}$ \cite{Decamp:1991uy}. Thus, we have 
\begin{eqnarray}
\sigma_{\nu N}\lesssim 4.1 \times 10^{-9} GeV^{-2}
\end{eqnarray} 
and, as the 4-fermion contribution to the cross section is
\begin{eqnarray*}
\sigma_{\nu N}=\left(\frac{\alpha_{S_0} v^2}{2 \Lambda^2}\right)^2 \frac{m_Z^2}{48\pi v^4}
\end{eqnarray*} 
the bound for the corresponding coupling is $\zeta_{S0} \leqslant 2.85\times 10^{-1}$.

On the other hand, we have the reaction $e^- e^+ \rightarrow N N$ with the bound obtained by LEP and shown in \eqref{eq:zNN}. Using the general expression for the cross section at the $Z$-pole 
\begin{eqnarray*}
\sigma_{N N}=\frac{12\pi}{m_Z^2}Br(Z\rightarrow e^+e^-)Br(Z\rightarrow N N)
\end{eqnarray*}
and in the low mass limit where $Br(N\rightarrow \nu(\bar \nu)\gamma)=1$ we have 
\begin{eqnarray}\label{eq:bNN}
\sigma_{NN}\lesssim 8.2 \times 10^{-9} GeV^{-2}
\end{eqnarray}
In the effective theory we are considering the operators that contribute to the Majorana neutrino pair production are the 4-fermion operators:
$ \mathcal{O}_{LN}$, $\mathcal{O}_{eNN}$ and $\mathcal{O}_{LNN}$ and the corresponding cross sections are
\begin{eqnarray}
\sigma^{NN}=\left(\frac{\alpha_{\mathcal{O}} v^2}{2 \Lambda^2}\right)^2 \frac{m_Z^2}{b_{\mathcal{O}}\pi v^4}
\end{eqnarray}
where $b_{eNN}=24$, $b_{LNN}=24$ and $b_{LN}=96$. Thus, using \eqref{eq:bNN} the most restrictive bound obtained is 
$\zeta_{4-f}\lesssim 2.63 \times 10^{-1}$.

In the case of 1-loop operators we have contributions to the $Z$-decay $Z \rightarrow \nu N$
\begin{eqnarray}
\Gamma(Z\rightarrow \nu N)= \left(\frac{\alpha^{1-loop} v^2}{2 \Lambda^2}\right)^2\frac{(c_W-s_W)^2}{6\pi}\frac{m_Z^3}{v^2}
\end{eqnarray}
and with the experimental bound for the Branching ratio \cite{Decamp:1991uy}
\begin{eqnarray}
Br(Z\rightarrow \nu N)=\frac{\Gamma(Z\rightarrow \nu N)}{\Gamma(Z\longrightarrow all)} \lesssim 2.7 \times 10^{-5}
\end{eqnarray}
we obtain $\Gamma(Z \rightarrow \nu N) \leq 6.7 \times 10^{-5}$, and thus
\begin{eqnarray}
\zeta_{1-loop}=\left(\frac{\alpha^{1-loop} v^2}{2 \Lambda^2}\right)^2 \lesssim 6.75 \times 10^{-4}
\end{eqnarray}

For the Lepton-Flavor-Violating processes e.g. $\mu \rightarrow e \gamma$, $\mu \rightarrow e e e$ and $\tau \rightarrow e e e$, induced by the quantum effect of the heavy neutrinos, we have very weak bounds for $m_N < m_W$ \cite{Antusch:2015mia, Tommasini:1995ii}.

\begin{figure}[htbp]
\begin{center}
\includegraphics[width=0.8\textwidth]{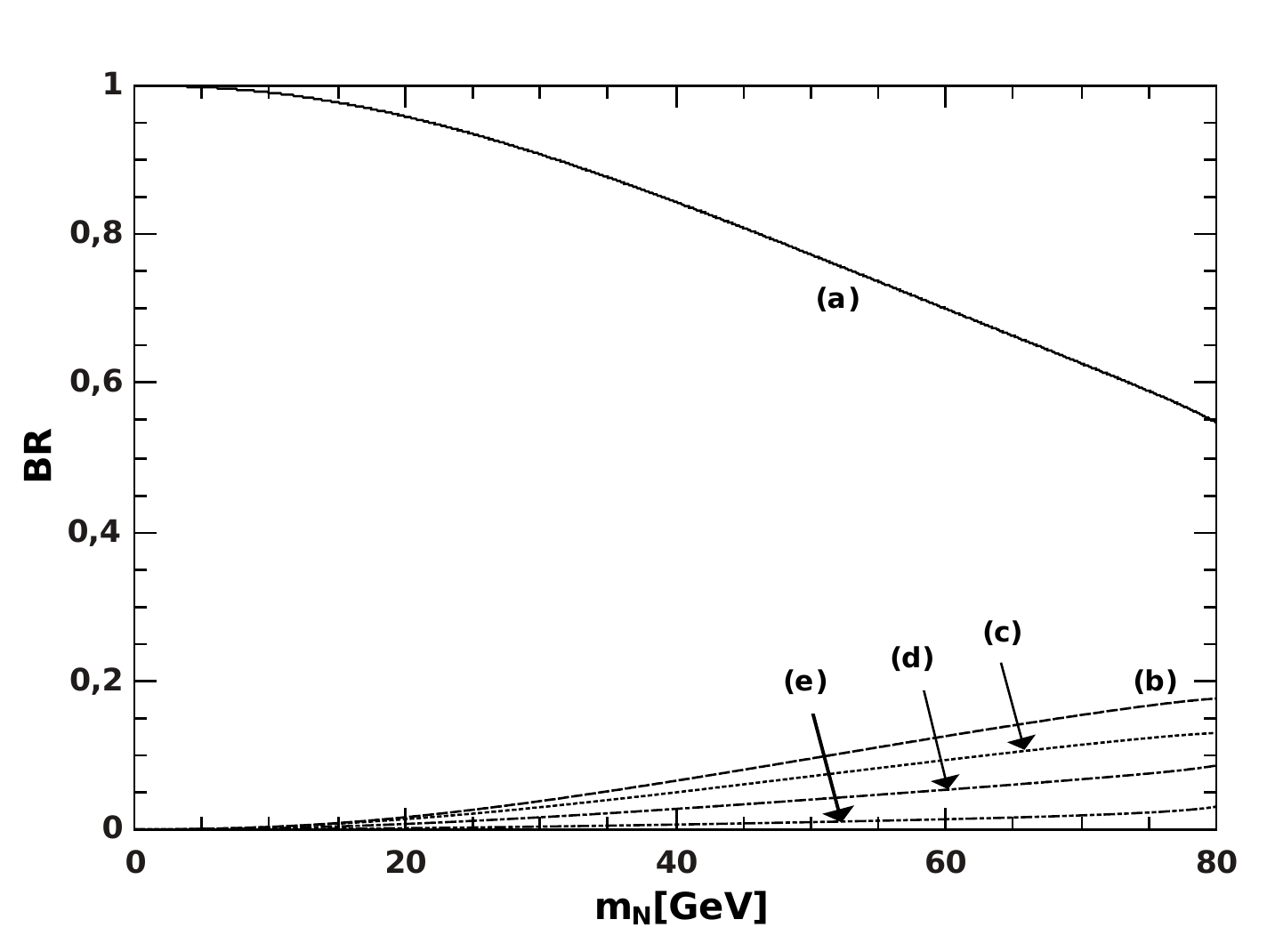} 
\hspace{4cm}
\caption{\label{fig:branching} Branching ratios for the Majorana neutrino decay channels in the low mass region. The expressions for the partial widths are presented in \cite{Duarte:2015iba}. The labels represent the decays {\bf{\scriptsize{(a)}}} $N\rightarrow \sum\limits_{i=1,3}\nu_i\, A$, {\bf{\scriptsize{(b)}}} $N\rightarrow \sum\limits_{i=1,3}d\bar d \nu_i (\bar \nu_i)$, {\bf{\scriptsize{(c)}}} $N\rightarrow \sum\limits_{i=1,3}u\bar u \nu_i (\bar \nu_i)$, {\bf{\scriptsize{(d)}}} $N\rightarrow leptons$ and {\bf{\scriptsize{(e)}}} $N \rightarrow \sum\limits_{i=1,3}u\bar d l_i$.}
\end{center}
\end{figure}

In the case of the heavy Majorana neutrino with effective interactions we are considering, the clear dominance of the neutrino plus photon decay channel, as we show in Fig.\ref{fig:branching} makes the beam dump and rare LNV experiments bounds inapplicable, as this decay mode to invisible particles is not considered in those analysis, and can considerably alter the number of events found for $N$ decays inside the detectors \cite{Deppisch:2015qwa, deGouvea:2015euy}. 

The results on the bounds for the different sets of coupling constants $\zeta_{\mathcal{O}}$ are summarized in table \ref{tab:bounds}. In consequence for the Majorana neutrino mass around $m_{\tau}$ we are safe from the most stringent bounds, and we will consider for simplicity the following set of limits for the operators of the respective sets:
\begin{eqnarray}\label{zetas-values}
\zeta_{SVB}&\lesssim & 7.6 \times 10^{-4}
\nonumber \\
\zeta_{4-f}&\lesssim & 2.7 \times 10^{-1}
\nonumber \\
\zeta_{1-loop}& \lesssim & 7. \times 10^{-4}
\end{eqnarray}
%

%%%%%%%%%%%%%%%%%%%%%%%%%%%%%%%%%%%%
\section{Neutrino propagation through the Earth}

\subsection{Relevant processes \label{procesos}}

In this section we study the different reactions taking place in the transport of tau-neutrinos in their journey through the Earth. We classify the produced 
effects as absorption and regeneration processes.

Absorption effects are all the processes that take out of the flux tau-neutrinos of energy $E$, and regeneration effects are those adding tau-neutrinos with energy $E$ to the flux. We must consider that beside ordinary neutrinos we have Majorana neutrinos and tau-leptons produced by the former when they pass through the Earth.

The standard interactions of ordinary neutrinos with the nucleons $\mathcal{N}$ forming the Earth are $\nu_{\tau} \mathcal{N} \rightarrow l^\pm X$ and $\nu_{\tau} \mathcal{N} \rightarrow \nu_{\tau} X$. Here the charged-current and neutral-current reactions contribute to the absorption effects of ordinary tau-neutrinos and neutral-current reactions contribute to the regeneration effects as we will discuss in the next section.

The production of Majorana neutrinos is driven by the collision of ordinary neutrinos with nucleons in the Earth, $\nu_{\tau} \mathcal{N}
\rightarrow N X$. This reaction absorbs ordinary tau-neutrinos producing Majorana neutrinos.

We also have to take into account the interaction of Majorana neutrinos with the nucleons forming the Earth, $N \mathcal{N} \rightarrow N X$, $N \mathcal{N} \rightarrow l^{\pm} X$ and $N \mathcal{N} \rightarrow \nu_{\tau} X$. In the same way that for the ordinary neutrinos, these reactions produce absorption effects of Majorana neutrinos, as well as regeneration effects for ordinary neutrinos and charged leptons.

Finally we have the Majorana neutrino dominant decay: $N \rightarrow \nu \; \gamma$ causing Majorana neutrino absorption and ordinary neutrino regeneration, and the standard $\tau$-decay leading to $\tau$ absorption and $\nu_{\tau}$ regeneration.

As we will discus in the next section, only some processes are relevant for the $\nu_{\tau}$ propagation in the Earth. We do not show explicitly expressions for the different cross-sections because it is a standard calculation. We prefer to show the results as plots for the interaction and decay lengths for the $SM$ scatterings $\nu_{\tau} \mathcal{N} \rightarrow l X$, $\nu_{\tau} \mathcal{N} \rightarrow \nu_{\tau} X$, the $\tau$-decay, the Majorana production process $\nu_{\tau} \mathcal{N} \rightarrow N X$, and the $N$-decay, as they give the relevant contributions to the transport equations. The dominant contribution is given by the 4-fermion operators $\mathcal{O}_{4-f}$.    

These lengths are defined in terms of the associated process cross-sections as:
\begin{eqnarray}\label{Lengths}
L_{\rm int}^{\rm tot, SM}(E)&=&\frac{1}{\langle\rho_n\rangle (\sigma^{\nu \mathcal{N}
\rightarrow l^{+} X}+\sigma^{\nu \mathcal{N}
\rightarrow \nu X})} \nonumber\\
L_{\rm int}^{\rm Maj}(E)&=&\frac{1}{\langle\rho_n\rangle \sigma^{\nu \mathcal{N}
\rightarrow N X}} \nonumber\\
L_{\rm decay}^{\rm \tau-decay}(E)&=&\frac{1}{\langle\rho_n\rangle \Sigma_{\tau}}
\nonumber\\
L_{decay}^{\rm N-decay}(E)&=&\frac{1}{\langle\rho_n\rangle \Sigma_{N}}.
\end{eqnarray}
Here $\langle\rho_n\rangle$ is the average number density along the column depth on the path with inclination $\theta$ with respect to the nadir direction as it will be shown in \eqref{eq:dens_prom}. The number density is defined as $\rho_n = N_A \rho$ where $N_A$ is the Avogadro constant and $\rho$ is the Earth mass density. The decay functions $\Sigma_{N~(\tau)}$ are defined in \eqref{eq:Sigmas_dec}.

\begin{figure*}[h]
\centering
\subfloat[{\scriptsize Interaction lengths: {\tiny{\bf{(1)}}} for the $SM$ $\nu_{\tau}$ scattering process ($L^{tot,SM}_{int}$), {\tiny{\bf{(2)}}} for the Majorana production process $\nu_{\tau} \mathcal{N} \rightarrow N X$  relevant for the $\nu_{\tau}$ absorption ($L^{Maj}_{int}$); and decay lengths: {\tiny{\bf{(3)}}} for the lepton $\tau$, $L^{\tau-decay}_{decay}$ and {\tiny{\bf{(4)}}} for the Majorana neutrino $N$, $L^{N-decay}_{decay}$.}]
{\label{fig:long1}\includegraphics[totalheight=6.5cm]{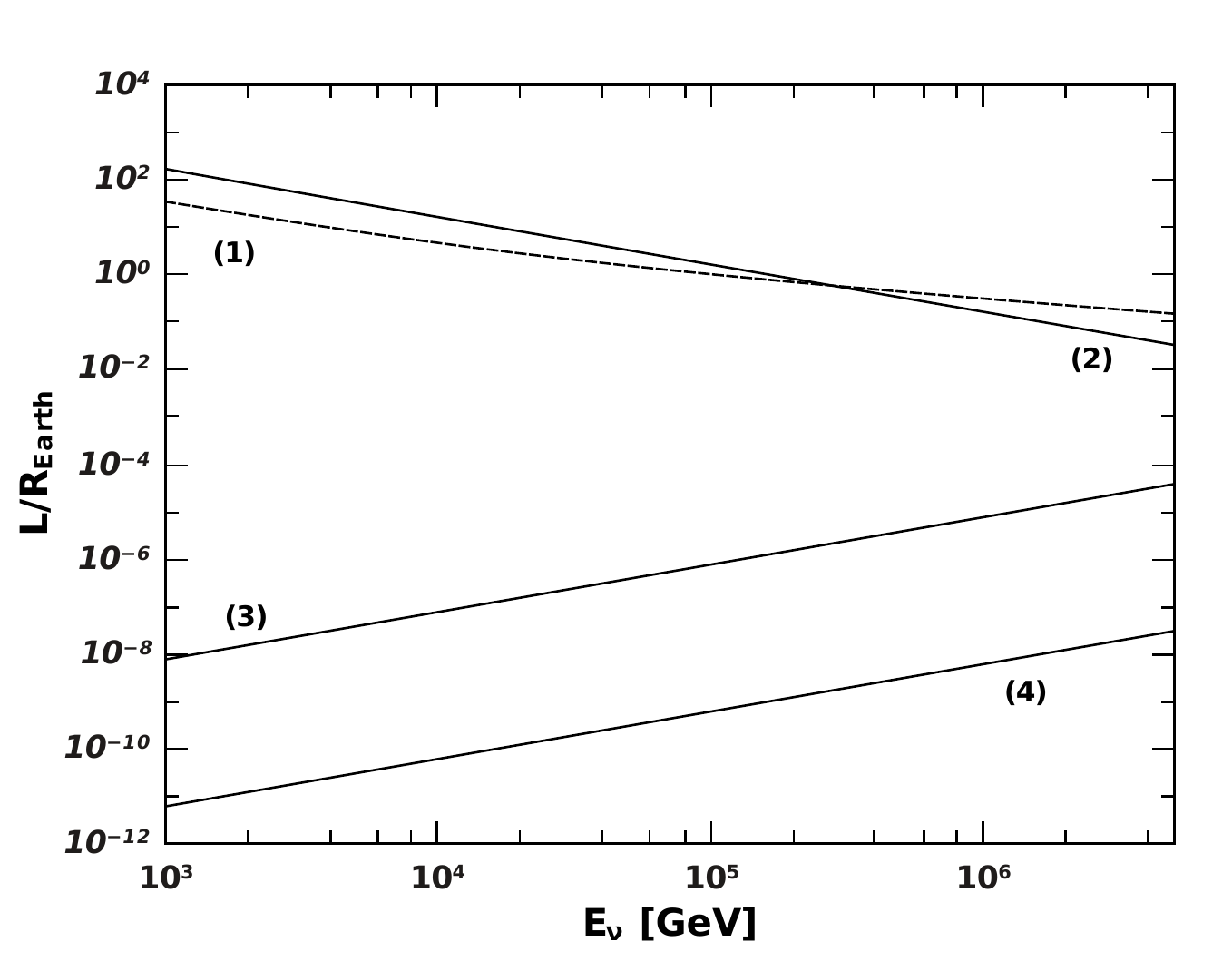}}~
\subfloat[{\scriptsize Interaction lengths for different Majorana neutrino processes: {\tiny{\bf{(1)}}} $N\mathcal{N}\rightarrow l X$, {\tiny{\bf{(2)}}} $N\mathcal{N}\rightarrow \nu_{\tau} X$, {\tiny{\bf{(3)}}} $\nu_{\tau} \mathcal{N}\rightarrow N X$ and {\tiny{\bf{(4)}}} $N\mathcal{N}\rightarrow N X$.}]{\label{fig:long2}\includegraphics[totalheight=6.5cm]{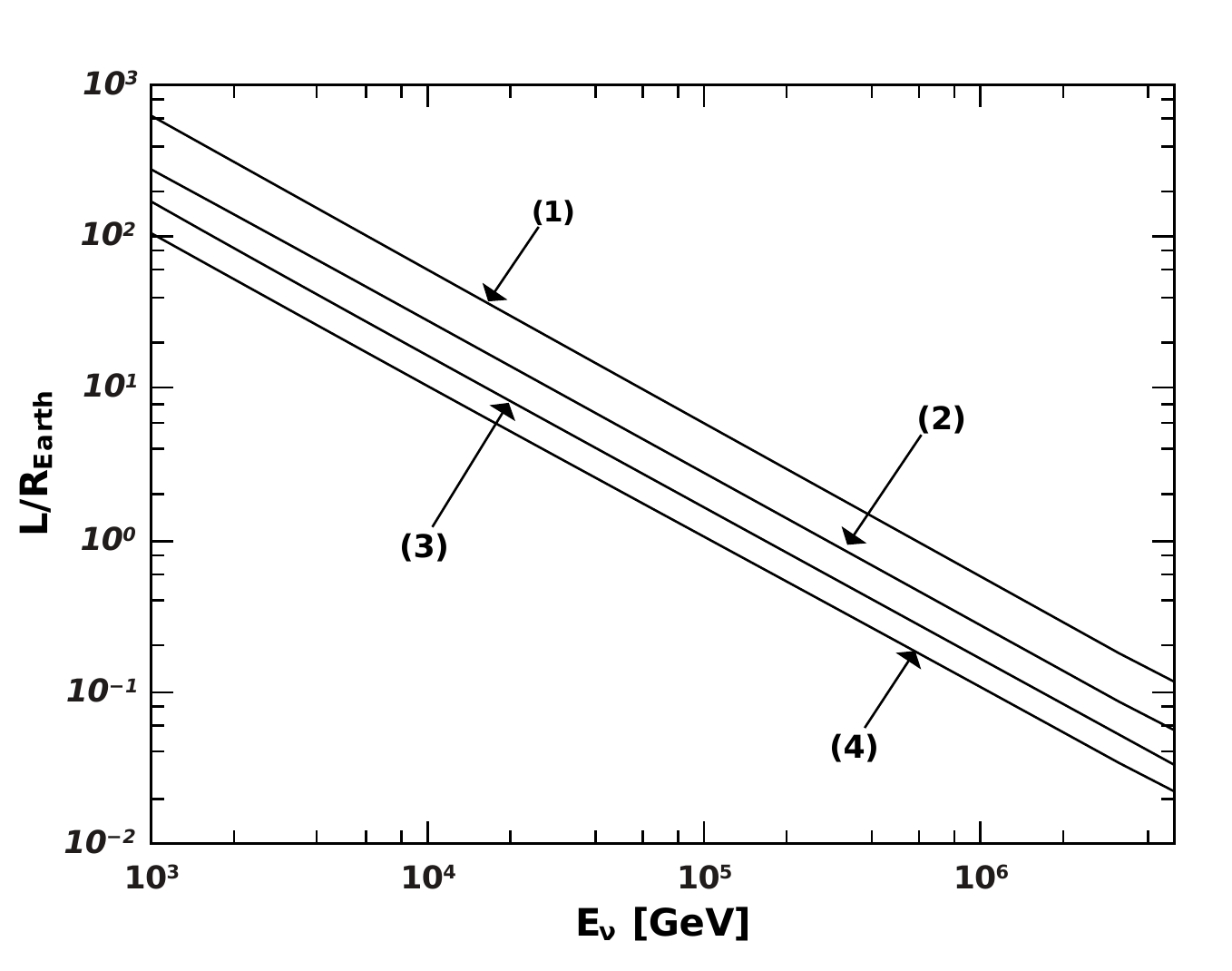}}
\caption{Comparison between the interaction and decay lengths for different neutrino energies as a fraction of the Earth radius,
 in a path in the nadir direction $\theta=0$, with the average density number calculated along this path.}
\label{fig:long}
\end{figure*}

In Fig.\ref{fig:long} we show the corresponding interaction and decay lenghts, along the nadir direction, as a fraction of the Earth radius, when the couplings take the upper values shown in Eq. \eqref{Lengths}.

In order to take into account the contribution of the $N$ decay to the regeneration through the $N \rightarrow \nu_{\tau}\gamma$ channel, we follow the
approach of Gaisser \cite{Gaisser:1990vg} which is developed in the appendix \ref{ap1}. We also take into account the regeneration effects coming from the $\tau$-decay according to expressions obtained by Gaisser \cite{Gaisser:1990vg} and also shown in \cite{Dutta:2000jv}. In Fig.\ref{fig:long2} we show for comparison the interaction length for the other involved interactions.

In the following section we will discuss the relative importance between the different contributions to the $\nu_{\tau}$ propagation.

\subsection{\bf Surviving neutrino flux \label{transporte}}
The neutrinos traveling through the Earth may suffer charged current (CC) and neutral current (NC) interactions with the nucleons in their path (see \cite{Nicolaidis:1996qu,Reynoso:2004dt} and references therein). Neutrino oscillation within the Earth can be neglected for energies higher than $1$ TeV \cite{Kwiecinski:1998yf}.

As we mentioned above, the change in neutrino flux $\Phi_{\nu_{\tau}}(E,\chi)$ as it traverses the Earth can be divided into two effects: absorption and regeneration. Absorption is a decrease in the neutrino flux of a given energy. In the $SM$ we have the total cross-section $\sigma^{\rm SM}_{\rm tot ~{\nu_{\tau}}}(E)=\sigma_{\rm CC}(E)+\sigma_{\rm NC}(E)$, which represents a probability of CC or NC standard $\tau$-neutrino interactions. When neutrinos pass through an amount of matter $d\chi=\rho_n(z) dz$ in a distance $dz$ along the neutrino beam path, where $\rho_n(z)$ is the Earth's number density, the change in the flux $\Phi_{\nu_{\tau}}(E,\chi)$ due only to absorption is proportional to $\Phi_{\nu_{\tau}}(E,\chi)$ and to the cross section:
\begin{eqnarray}\label{asu0}
\frac{d\Phi_{\nu_{\tau}}(E,\chi)}{\hspace{-0.6cm}d\chi}=-\sigma_{\rm tot ~{\nu_{\tau}}}^{\rm SM}(E)
\Phi_{\nu_{\tau}}(E,\chi).
\end{eqnarray}
Here $\chi(z)$ is the amount of material found up to a depth $z$, that is,
\begin{eqnarray}\label{chi}
\chi(z) = \int^z_0 dz' \rho_n(z'),
\end{eqnarray}
where the number density is the Avogadro's constant times the density, $\rho_n(z')=N_A \rho(z')$. 

In order to consider the complete transport effect for UHE neutrinos, we have to add to (\ref{asu0}) the effect of regeneration, which accounts for the possibility that neutrinos of energies $E'>E$ may end up with energy $E$ due to NC interactions with the nucleons, adding neutrinos to the flux of energy $E$. Then, the $SM$ transport equation for neutrinos reads
\begin{eqnarray}
\frac{\partial\Phi_{\nu_{\tau}}(E,\chi)}{\partial\chi}=-\sigma^{\rm SM}_{\rm tot~{\nu_{\tau}}}(E)\Phi_{\nu_{\tau}}(E,\chi)+
\int\limits_0^{1} \frac{dy}{(1-y)} \Phi_{\nu_{\tau}}\left(E/(1-y),\chi\right)
\frac{d\sigma^{\nu_{\tau} \mathcal{N} \rightarrow \nu_{\tau} X}}{dy}(E,y)
\end{eqnarray}
Here the usual change of variables $y= (E'-E)/ E'$ has been made.

On the other hand, taking into account the sterile Majorana neutrinos production and decay processes, we have new contributions to the
absorption and regeneration effects on the $\nu_{\tau}$-flux. Moreover, as it is well known, if we consider the $\nu_{\tau}$ transport, it is important to take into account the transport of the $\tau$-lepton which regenerates the $\nu_{\tau}$ by $\tau$-decay. In the same way, the Majorana neutrino decay can regenerate the $\nu_{\tau}$ flux. So in principle one needs to simultaneously solve a system of three coupled integro-differential equations:

\begin{eqnarray} \label{transporte_nu}
&&\frac{\partial\Phi_{\nu_{\tau}}(E,\chi)}{\partial\chi}=-\sigma_{\nu_{\tau}}^{\rm t}(E)
\Phi_{\nu_{\tau}}(E,\chi)+\sigma_{\nu_{\tau}}^{\rm t}(E)\int_0^1 \frac{dy}{(1-y)}
\Phi_{\nu_{\tau}}(E_y,\chi)K^{\rm NC}_{\nu_{\tau}}(E,y) \nonumber\\
&+& \sigma^{\rm t}_{\tau}(E) \int_0^1 \frac{dy}{(1-y)} \Phi_{\tau}(E_y,\chi) K_{\tau}^{\rm cc}(E,y) 
+ \sigma^{\rm t}_{N}(E) \int_0^1 \frac{dy}{(1-y)} \Phi_{N}(E_y,\chi) K_{N}^{\rm N\mathcal{N} \rightarrow \nu_{\tau} X}(E,y) 
\nonumber\\
&+& \Sigma_{\tau}(E) \int_0^1 \frac{dy}{(1-y)} \Phi_{\tau}(E_y,\chi) K^{\rm dec}_{\tau}(E,y) 
+ \Sigma_{N}(E) \int_0^1 \frac{dy}{(1-y)} \Phi_{N}(E_y,\chi) K^{\rm dec}_{N}(E,y)
\end{eqnarray}

\begin{eqnarray}\label{transporte_tau}
&&\frac{\partial\Phi_{\tau}(E,\chi)}{\partial\chi}=-\sigma_{\tau}^{\rm t}(E)
\Phi_{\tau}(E,\chi)-\Sigma_{\tau} \Phi_{\tau}(E,\chi)+\frac{d}{dE}(E\beta(E)\Phi_{\tau}(E))+
\nonumber\\
&& \sigma_{\tau}^{\rm t}(E)\int_0^1 \frac{dy}{(1-y)}
\Phi_{\tau}(E_y,\chi)K^{\rm NC}_{\tau}(E,y) 
+\sigma_{\nu_{\tau}}(E) \int_0^1 \frac{dy}{(1-y)} \Phi_{\nu_{\tau}}(E_y,\chi) K_{\nu_{\tau}}^{\rm cc}(E,y) 
\nonumber\\
&&+ \sigma_{N}(E) \int_0^1 \frac{dy}{(1-y)} \Phi_{N}(E_y,\chi) K_N^{\rm N\mathcal{N}\rightarrow\tau X}(E,y) 
\end{eqnarray}

\begin{eqnarray}\label{transporte_N}
&&\frac{\partial\Phi_{N}(E,\chi)}{\partial\chi}=-\sigma^{\rm t}_N(E) \Phi_N(E,\chi)
-\Sigma_{N}(E) \Phi_{N}(E,\chi)
\nonumber\\
&+&\sigma^{\rm t}_{\nu_{\tau}}(E) \int_0^1
\frac{dy}{(1-y)} \Phi_{\nu_{\tau}}(E_y,\chi) K_{\nu_{\tau}}^{\rm \nu_{\tau} \mathcal{N}\rightarrow N X}(E,y)
+\sigma^{\rm t}_{\tau}(E) \int_0^1
\frac{dy}{(1-y)} \Phi_{\tau}(E_y,\chi) K_{\tau}^{\rm \tau \mathcal{N}\rightarrow N X}(E,y)
\nonumber\\
&+& \sigma^{\rm t}_{N}(E) \int_0^1
\frac{dy}{(1-y)} \Phi_{N}(E_y,\chi) K_N^{\rm N \mathcal{N}\rightarrow N X}(E,y)
\end{eqnarray}

In \eqref{transporte_nu} the right hand side terms correspond to the absorption, neutral-current regeneration, charged current 
regeneration by $\tau$ interaction, regeneration by Majorana neutrino interaction, and regeneration by $\tau$ and $N$ decays. 

The Eq.\eqref{transporte_tau} corresponds to the $\tau$ transport with absorption by interaction and decay, and regeneration by $\tau$ neutral current, $\nu_{\tau}$ charged current and by the interaction of Majorana neutrino through the $N \mathcal{N}\rightarrow \tau X$ reaction. The third term represents the energy loss due to electromagnetic interactions. 

In the Eq.\eqref{transporte_N} we have absorption terms by interaction and by $N$ decay. The other terms represent the $N$-flux regeneration by the $\nu_{\tau}$, $\tau$ and $N$ interactions with nucleons.

In the above equations the cross-sections in the absorption by interaction terms are:
\begin{eqnarray}\label{eq:sigmas}
 \sigma^{\rm t}_{\nu_{\tau}}(E)&=&\sigma^{\rm SM}_{\rm tot ~ \nu_{\tau}}(E)+\sigma^{\nu_{\tau} \mathcal{N} \rightarrow N X}_{\nu_{\tau}}(E) 
\end{eqnarray}
and $\sigma^{\rm t}_{N}(E)$  and $\sigma^{\rm t}_{\tau}(E)$ include all the Majorana neutrino interactions. 

The different cross-section regeneration kernels are:
\begin{eqnarray} \label{eq:kapa_s}
K^{\rm NC}_{\nu_{\tau}}(E,y)&=&\frac{1}{\sigma^{\rm t}_{\nu_{\tau}}(E)}\frac{d\sigma^{\nu_{\tau} \mathcal{N} \rightarrow \nu_{\tau} X}(E_y,y)}{dy}
\;\;\mbox{,}\;\; 
K^{\rm CC}_{\tau}(E,y)=\frac{1}{\sigma^{\rm t}_{\tau}(E)}\frac{d\sigma^{\tau \mathcal{N} \rightarrow \nu_{\tau} X}(E_y,y)}{dy}
\nonumber\\ %
K_N^{\rm N\mathcal{N}\rightarrow \nu_{\tau} X}(E,y)&=&\frac{1}{\sigma^{\rm t}_{N}(E)}\frac{d\sigma^{N\mathcal{N}\rightarrow \nu_{\tau} X}(E_y,y)}{dy}
 \nonumber\\%
K^{\rm NC}_{\tau}(E,y)&=&\frac{1}{\sigma^{\rm t}_{\tau}(E)}\frac{d\sigma^{\tau \mathcal{N} \rightarrow \tau X}(E_y,y)}{dy}
\;\;\mbox{,}\;\; 
K^{\rm CC}_{\nu_{\tau}}(E,y)=\frac{1}{\sigma^{\rm t}_{\nu_{\tau}}(E)}\frac{d\sigma^{\nu_{\tau} \mathcal{N} \rightarrow \tau X}(E_y,y)}{dy}
\nonumber\\ %
K^{\rm N \mathcal{N}\rightarrow \tau X}_{N}(E,y)&=&\frac{1}{\sigma^{\rm t}_{N}(E)}\frac{d\sigma^{N \mathcal{N} \rightarrow \tau X}(E_y,y)}{dy}\;\;\nonumber\\ %
K^{\rm \nu_{\tau} \mathcal{N}\rightarrow N X}_{\nu_{\tau}}(E,y)&=&\frac{1}{\sigma^{\rm t}_{\nu_{\tau}}(E)}\frac{d\sigma^{\nu_{\tau} \mathcal{N} \rightarrow N X}(E_y,y)}{dy}
\;\;\mbox{,}\;\;    
K^{\rm \tau \mathcal{N}\rightarrow N X}_{\tau}(E,y)=\frac{1}{\sigma^{\rm t}_{\tau}(E)}\frac{d\sigma^{\tau \mathcal{N} \rightarrow N X}(E_y,y)}{dy} 
\nonumber\\ 
K^{\rm N \mathcal{N}\rightarrow N X}_N(E,y)&=&\frac{1}{\sigma^{\rm t}_{N}(E)}\frac{d\sigma^{N \mathcal{N} \rightarrow N X}(E_y,y)}{dy} 
\nonumber\\ 
\end{eqnarray}
with $E_{y}=E'=E/(1-y)$.

The decay kernels for the Majorana neutrino $N$ or $\tau$-lepton are calculated in the appendix \ref{ap1} and in \cite{Dutta:2000jv} respectively:
\begin{eqnarray}\label{eq:K-dec}
 \nonumber\\ %
K^{\rm dec}_{N(\tau)}(E,y)&=&(1-y) \frac{dn_{N(\tau)}(1-y)}{dy}
\end{eqnarray}
and the decay-length functions are
\begin{eqnarray}\label{eq:Sigmas_dec}
 \Sigma_{N(\tau)}(E)=\left( \frac{E}{m_{N(\tau)}} \langle \rho_n \rangle T_{N(\tau)} \right)^{-1}
\end{eqnarray}
where
$T_{N(\tau)}=(\Gamma_{rest~N(\tau)}^{tot})^{-1}$ is the $N (\tau)$ lifetime in its rest frame,
with
\begin{eqnarray} \label{photon_neutrino}
\Gamma_{rest~N}^{tot}=\frac{1}{4 \pi}\left[\sum_{i=1}^3 \left(\alpha^i_{NB}  c_W +  \alpha^i_{NW} s_W \right)^2 \right]
\frac{v^2}{m_N} \left( \frac{m_N}{\Lambda} \right)^4
\end{eqnarray}
as for the $N$ low mass range the dominant decay is $N \rightarrow \nu \gamma$.

Some of the terms in the equations above can be neglected in the considered energy range. The $\tau$ and $N$ interactions are neglected against their decays. The $\tau$ interactions begin to be dominant at an energy around $E_{\nu_{\tau}}=10^8$G$e$V \cite{Dutta:2000jv}. For the $N$ interactions, the contributions of the different processes are compared in Fig.\ref{fig:long} as a plot for the ratio between the interaction and decay lengths and the Earth radius. We neglect the regeneration terms coming from the $\tau$ and $N$ interactions, which are proportional to the $\tau$ and $N$ flux, in comparison with those due to NC interactions of $\nu_{\tau}$ and those originated in the $\tau$ and $N$ decay. In these conditions, for the $\nu_{\tau}$ transport equation \eqref{transporte_nu} we take into account the absorption and neutral-current regeneration terms and the regeneration by $\tau$ and $N$-decay. For the $\tau$-transport equation, we consider absorption by $\tau$-decay and regeneration by $\nu_{\tau}$ scattering by nucleons, which is the source of $\tau$ leptons appearance. In the case of the Majorana neutrino transport equation, absorption by $N$-decay and also regeneration by $\nu_{\tau}$ scattering by nucleons are included, the last process being the source for the $N$-flux.  

Finally, the equations we need to solve are:
\begin{eqnarray} \label{final_transporte_nu} 
\frac{\partial\Phi_{\nu_{\tau}}(E,\chi)}{\partial\chi}&=&-\sigma_{\nu_{\tau}}^{\rm t}(E)
\Phi_{\nu_{\tau}}(E,\chi) + \sigma_{\nu_{\tau}}^{\rm t}(E)\int_0^1 \frac{dy}{(1-y)}
\Phi_{\nu_{\tau}}(E_y,\chi)K^{\rm NC}_{\nu_{\tau}}(E,y)
\nonumber \\ 
&+&
\Sigma_{\tau}(E) \int_0^1 \frac{dy}{(1-y)} \Phi_{\tau}(E_y,\chi) K^{\rm dec}_{\tau}(E,y)
\nonumber \\ 
&+& \Sigma_{N}(E) \int_0^1 \frac{dy}{(1-y)} \Phi_{N}(E_y,\chi) K^{\rm dec}_{N}(E,y) 
\end{eqnarray}

\begin{eqnarray}\label{final_transporte_tau}
\frac{\partial\Phi_{\tau}(E,\chi)}{\partial\chi}=-\Sigma_{\tau}(E) \Phi_{\tau}(E,\chi)
+\sigma^{\rm t}_{\nu_{\tau}}(E) \int_0^1 \frac{dy}{(1-y)} \Phi_{\nu_{\tau}}(E_y,\chi) K_{\nu_{\tau}}^{\rm CC}(E,y) 
\end{eqnarray}

\begin{eqnarray}\label{final_transporte_N}
\frac{\partial\Phi_{N}(E,\chi)}{\partial\chi}=-\Sigma_{N}(E) \Phi_{N}(E,\chi)
+\sigma^{\rm t}_{\nu_{\tau}}(E) \int_0^1
\frac{dy}{(1-y)} \Phi_{\nu_{\tau}}(E_y,\chi) K^{\rm \nu_{\tau} \mathcal{N}\rightarrow N X}_{N}(E,y)
\end{eqnarray}

The system of transport equations (Eqs.\eqref{final_transporte_nu}- \eqref{final_transporte_N}) must be solved with the initial conditions
$\Phi_{\nu_{\tau}}(E,\chi=0)=\Phi_{\nu_{\tau}}^0(E,\theta)$, $\Phi_{\tau}(E,\chi=0)=0$ and $\Phi_{N}(E,\chi=0)=0$, where $\Phi_{\nu_{\tau}}^0(E,\theta)$ is an initial neutrino flux.

\begin{figure}
\begin{center}
\includegraphics[width=0.6\textwidth]{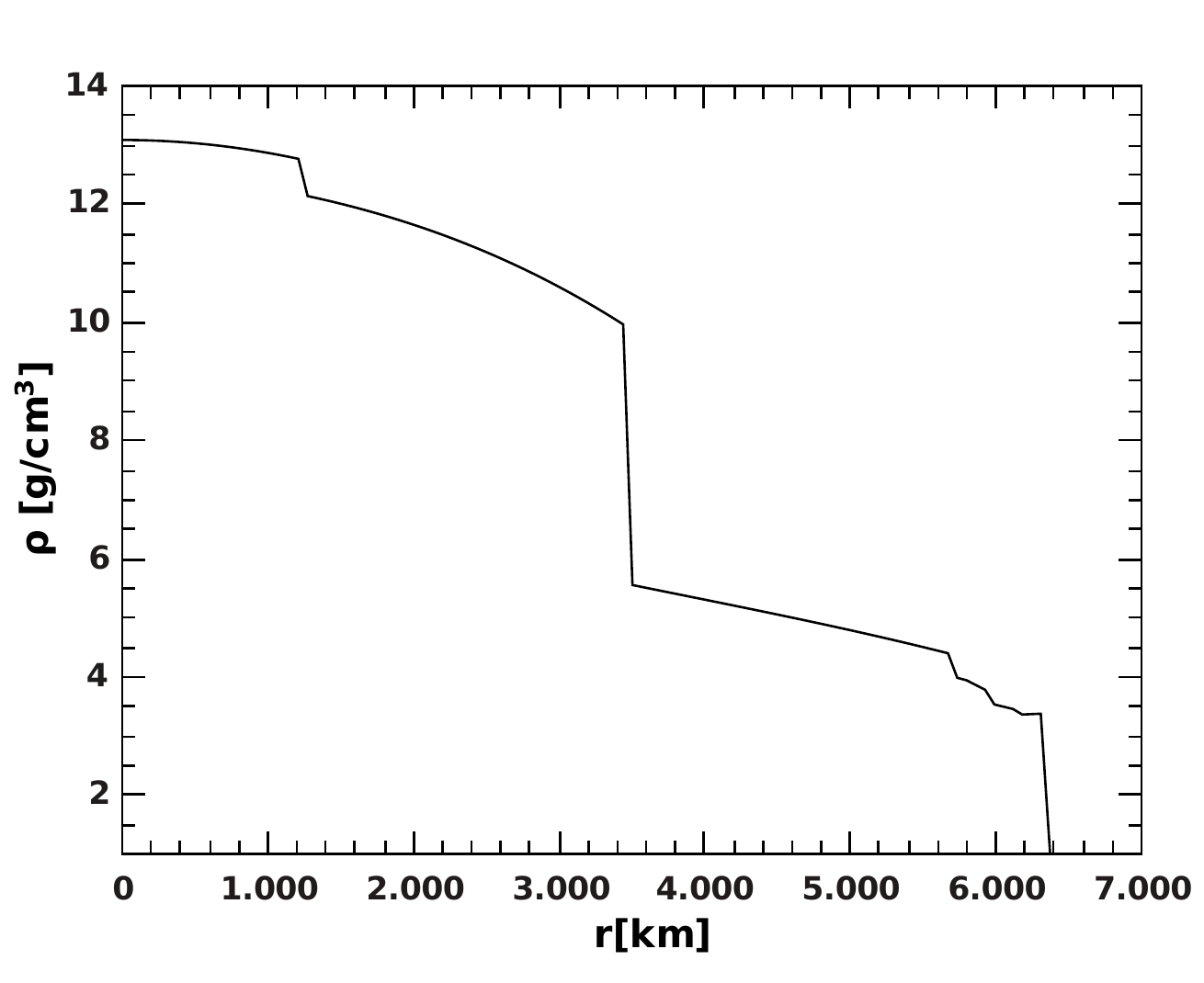}
%\hspace{4cm}
 \caption{\label{fig:densidad} Earth density as given by the PREM \citep{DZIEWONSKI1981297}}
\end{center}
\end{figure}

Taking the column depth on the path with inclination $\theta$ respective to the nadir direction taken from the down-going normal to the neutrino telescope as $\mathcal{T}(\theta)$:
\begin{eqnarray}\label{Tdensidad}
\mathcal{T}(\theta)=\chi(2 R \cos\theta)=\int\limits_0^{2 R\cos\theta} \rho_n(z) dz,
\end{eqnarray}
with $R$ as the Earth radius, we define $\langle \rho_n \rangle$ as the average number density along the column
\begin{eqnarray}\label{eq:dens_prom}
\langle\rho_n(\theta)\rangle=\frac{\mathcal{T}(\theta)}{2 R \cos\theta}.
\end{eqnarray}
The Earth density is given by the Preliminary Reference Earth Model (PREM) \cite{DZIEWONSKI1981297}. In Fig.\ref{fig:densidad} we present the Earth density profile. 

In accordance with \cite{Jain:2002kz}, and following the treatment made in \cite{Iyer:1999wu, Dutta:2000jv, Reynoso:2013wea}, 
we solve Eqs. \eqref{final_transporte_tau} and \eqref{final_transporte_N} considering the terms dependent of the $\nu_{\tau}$ flux $\Phi_{\nu_{\tau}}(E,\chi)$ as non-homogeneities, and replace those solutions in \eqref{final_transporte_nu}, dividing by $\Phi_{\nu_{\tau}}(E, \chi)$. Then we make an approximation, taking the fluxes quotient \eqref{flux-quotients} as the ones solving the corresponding homogeneous equations \cite{Jain:2002kz}.  
Finally, we write the solution for the surviving $\tau$-neutrino flux traversing a path of length $\mathcal{T}(\theta)$ through the Earth in terms of $\sigma_{\rm eff}(E,\mathcal{T}(\theta))$ as:
\begin{eqnarray}
\Phi_{\nu_{\tau}}(E,\mathcal{T}(\theta))=\Phi_{\nu_{\tau}}(E,0)
\exp[-\sigma_{\rm eff}(E,\mathcal{T}(\theta)) \mathcal{T}(\theta)],
\end{eqnarray}
with
\begin{eqnarray}
&&\sigma_{\rm eff}(E,\mathcal{T}(\theta))=\sigma^t_{\nu_{\tau}}(E)-\sigma^t_{\nu_{\tau}}(E) \int_0^1
dy \xi(E,y) K^{\rm NC}_{\nu_{\tau}}(E,y) \mathcal{D}(E_y,E) 
\nonumber \\
&-& \int_0^1\int_0^1 dy dy^{\prime} \xi(E,y) \xi(E_y,y^{\prime}) 
\frac{\Sigma_{\tau}(E) K_{\tau}^{\rm dec}(E,y) \sigma^t_{\nu_{\tau}}(E_y)
K^{\rm CC}_{\tau}(E_y,y^{\prime})}{\Delta_{\tau}(E_y,E_{yy^{\prime}})}
%\nonumber \\ &&
 (\mathcal{D}(E_{yy^{\prime}},E)-\mathcal{D}_{\tau}(E_{y},E))
\nonumber \\
&-&\int_0^1\int_0^1 dy dy^{\prime} \xi(E,y) \xi(E_y,y^{\prime}) 
\frac{\Sigma_{N}(E) K_{N}^{\rm dec}(E,y) \sigma^t_{\nu_{\tau}}(E_y)
K_{\nu_{\tau}}^{\rm \nu_{\tau}\mathcal{N}\rightarrow N X}(E_y,y^{\prime}) }{\Delta_{N}(E_y,E_{y y^{\prime}})} 
%\nonumber \\ && 
(\mathcal{D}(E_{yy^{\prime}},E)-\mathcal{D}_{N}(E_{y},E))
\nonumber\\
\end{eqnarray}
Here $E_{yy^{\prime}}=E/((1-y)(1-y^{\prime}))$ and the flux quotients are

\begin{eqnarray}\label{flux-quotients}
 \xi(E,y)=\frac{1}{(1-y)}\frac{\Phi^{0}_{\nu_{\tau}}(E_y)}{\Phi^{0}_{\nu_{\tau}}(E)} \; , ~~~~
 \xi(E_y,y^{\prime})=\frac{1}{(1-y^{\prime})}\frac{\Phi^{0}_{\nu_{\tau}}(E_{y y^{\prime}})}{\Phi^{0}_{\nu_{\tau}}(E)}
\end{eqnarray}
with
\begin{eqnarray*}
\mathcal{D}(E_1,E_2)&=&\frac{[1-\exp(-\Delta(E_1,E_2)
\mathcal{T}(\theta))]}{\Delta(E_1,E_2) \mathcal{T}(\theta)} \; , ~~~~
\Delta(E_1,E_2)=\sigma^t_{\nu_{\tau}}(E_1)-\sigma^t_{\nu_{\tau}}(E_2)
\nonumber \\
\mathcal{D}_{\tau}(E_1,E_2)&=&\frac{[1-\exp(-\Delta_{\tau}(E_1,E_2)
\mathcal{T}(\theta))]}{\Delta_{\tau}(E_1,E_2) \mathcal{T}(\theta)} \; ,~~~
\Delta_{\tau}(E_1,E_2)=\Sigma_{\tau}(E_1)-\sigma^t_{\nu_{\tau}}(E_2)
\nonumber \\
\mathcal{D}_{N}(E_1,E_2)&=&\frac{[1-\exp(-\Delta_N(E_1,E_2)
\mathcal{T}(\theta))]}{\Delta_N(E_1,E_2) \mathcal{T}(\theta)} \; ,~~~
\Delta_{N}(E_1,E_2)=\Sigma_{N}(E_1)-\sigma^t_{\nu_{\tau}}(E_2)
\end{eqnarray*}
\section{Numerical results \label{resultados}}

In this section we present our results assuming a Majorana neutrino contribution with $m_N \thicksim m_{\tau}$. In order to obtain numerical results for the surviving $\nu_{\tau}$ flux including Majorana neutrino effects, we consider a particular choice for the effective coupling constants $\zeta$, with the upper values presented in \eqref{zetas-values}. Also, for the initial $\nu_{\tau}$ flux we have considered the best fit of IceCube 
$\Phi^{0}_{\nu_{\tau}}=2.3 \times 10^{-18} (E/100TeV)^{-2.6} ~GeV^{-1} ~cm^{-2}~ s^{-1}~sr^{-1}$ \cite{Aartsen:2015ivb,Aartsen:2015knd}.

The idea is to see whether the effect of the Majorana neutrino modifies the $\nu_{\tau}$ surviving flux and to what extent it should be distinguishable from the standard surviving flux when a detection is performed in a neutrino telescope, which will clearly depend on the uncertainty involved.

\begin{figure*}[h]
\centering
\subfloat[{\scriptsize The quotient $R^{SM}(\theta,E_{\nu_{\tau}})=\Phi_{\nu_{\tau}}(\theta,E)/\Phi_{\nu_{\tau}}^{SM}(\theta,E)$ for different nadir angles. $\Phi_{\nu_{\tau}}(E,\theta)$ include SM and Majorana neutrino effects. }]
{\label{fig:rsm}\includegraphics[totalheight=6.5cm]{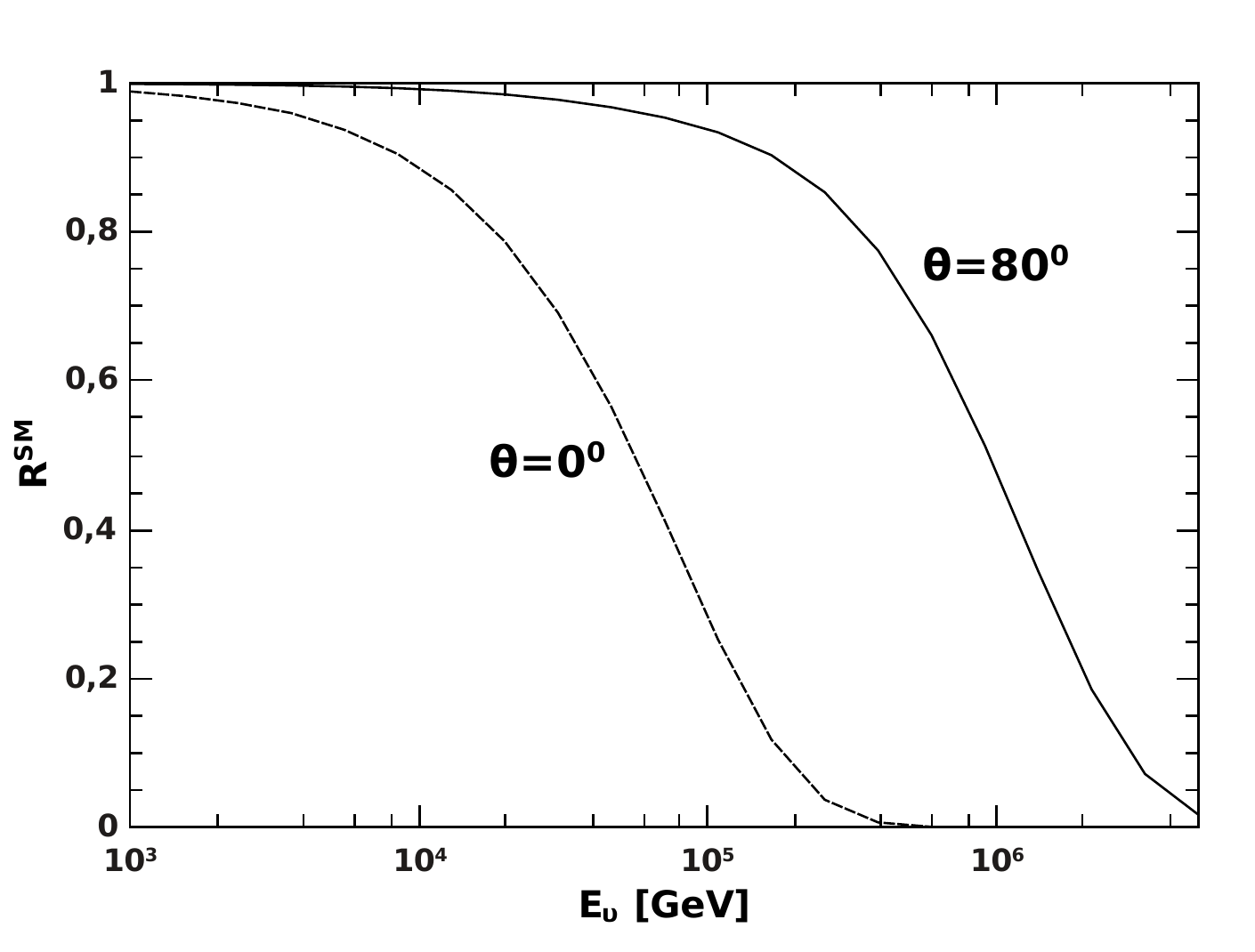}}~
\subfloat[{\scriptsize  The quotient $R^0(\theta,E_{\nu_{\tau}})=\Phi_{\nu_{\tau}}(\theta,E)/\Phi_{\nu_{\tau}}^0(E)$ for different nadir angles. For the curve labeled {\bf a} $\Phi_{\nu_{\tau}}(E,\theta)$ is the SM flux after traversing the Earth. The curve labeled {\bf b} also includes the Majorana neutrino effects. $\Phi^0_{\nu_{\tau}}(E,\theta)$ is the initial flux arriving to the Earth's surface.}]{\label{fig:r0}\includegraphics[totalheight=6.5cm]{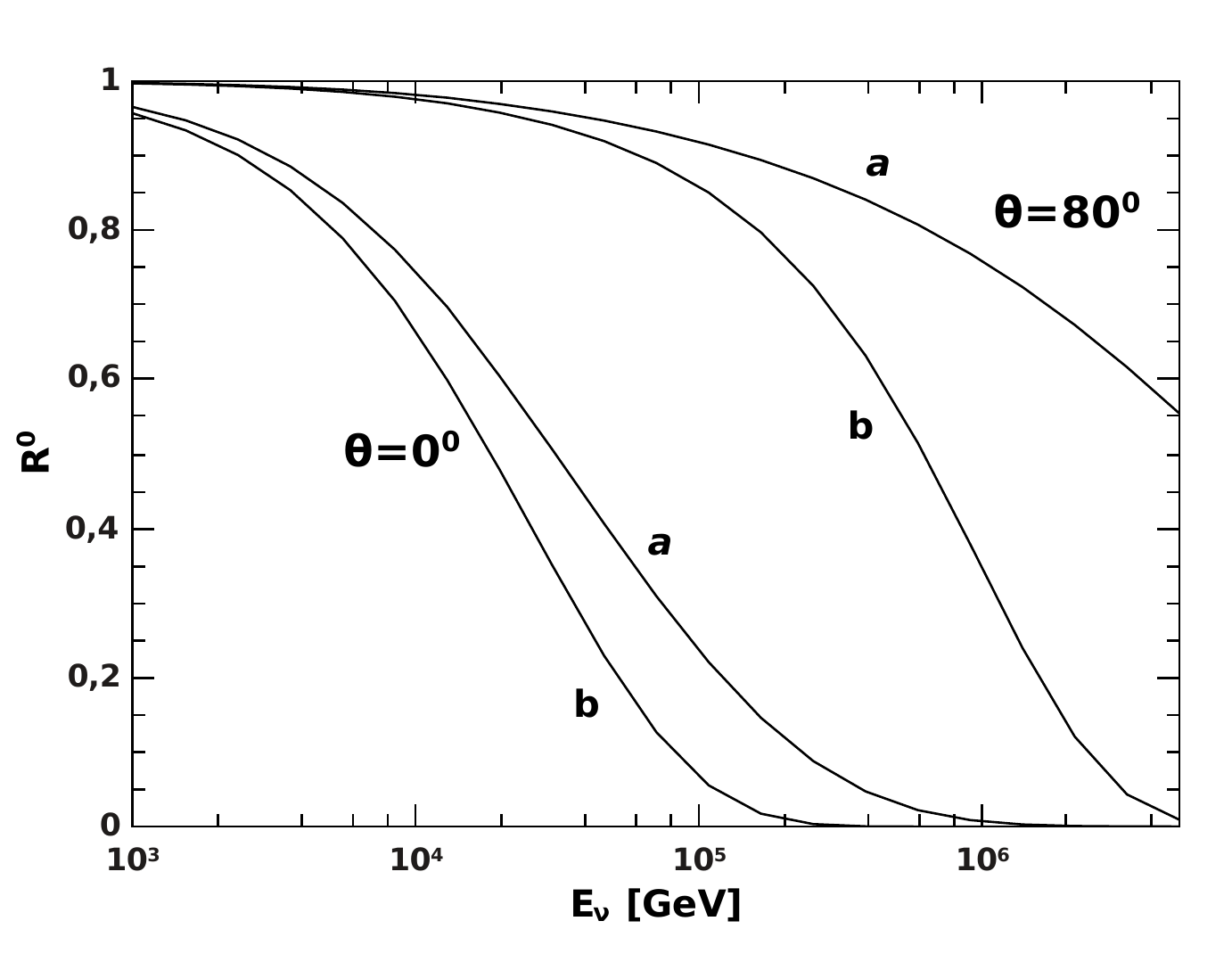}}
\caption{Ratio between $\nu_{\tau}$ fluxes for different angles.}\label{fig:r}
\end{figure*}

First, in Fig.\ref{fig:r}, we compare the surviving $\nu_{\tau}$ flux with Majorana neutrino effects, taking into account both absorption and regeneration, with the $SM$ prediction. In Fig.\ref{fig:rsm}, we show the comparison with the $SM$, showing the quotient $R^{SM}(\theta,E_{\nu_{\tau}})=\Phi_{\nu_{\tau}}/\Phi_{\nu_{\tau}}^{SM}$ for different nadir angles $\theta$. We also include a figure (Fig.\ref{fig:r0}) with the quotient between the surviving flux and the initial flux, $R^0(\theta,E_{\nu_{\tau}})=\Phi_{\nu_{\tau}}/\Phi_{\nu_{\tau}}^{0}$.

\begin{figure}
\begin{center}
\includegraphics[width=0.6\textwidth]{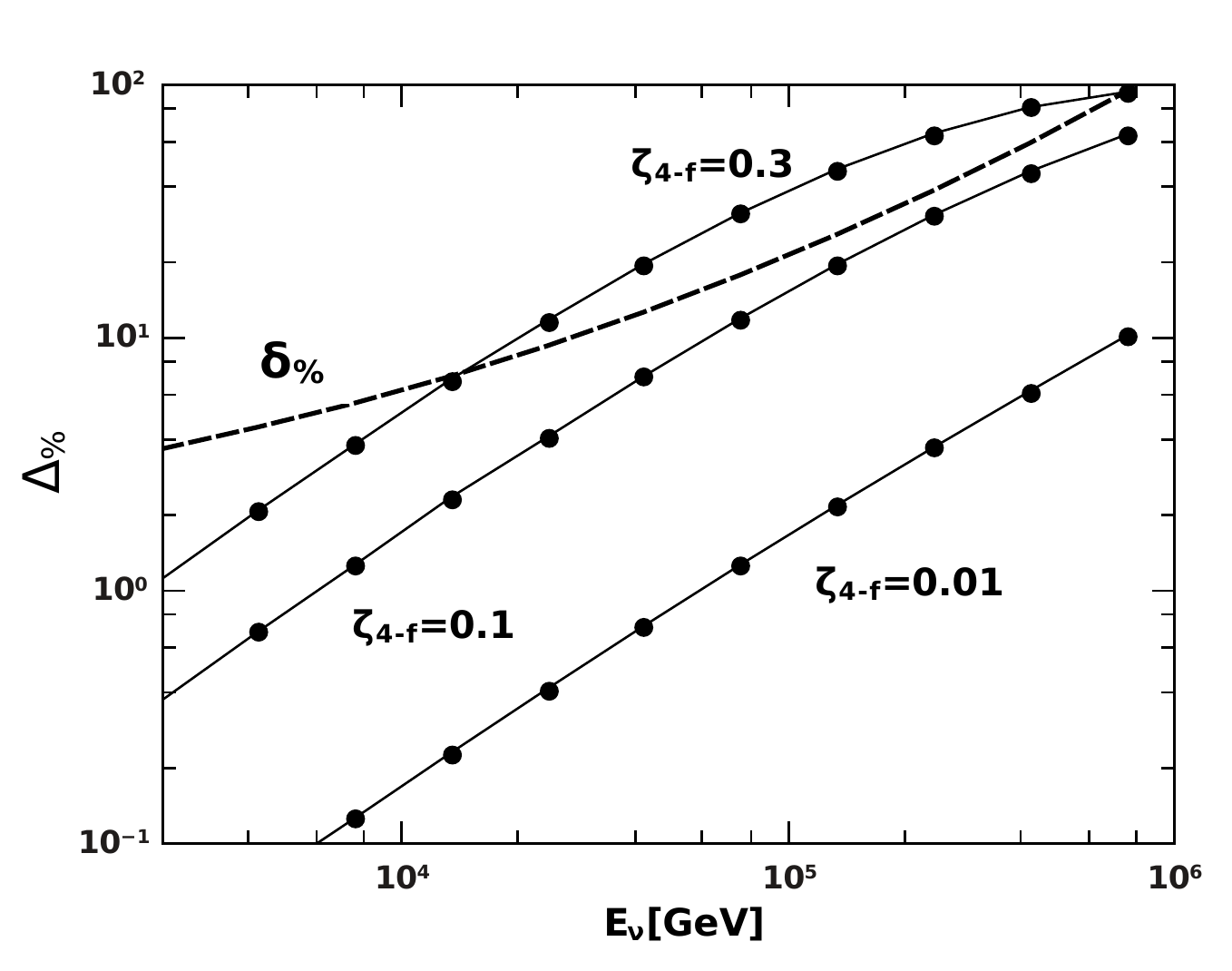}
%\hspace{4cm}
 \caption{\label{fig:delta} Percentage deviation of the number of events ($\Delta_{\%}$) and comparison with $SM$ statistical error ($\delta_{\%}$)
 for different values of the 4-fermion couplings $\zeta_{4-f}$. The solid circles indicate the center of the considered energy bin.}
\end{center}
\end{figure}

In order to calculate the capability of IceCube to detect the effects of Majorana neutrinos physics, we have considered an approximated number of events as given by
\begin{eqnarray}
 N=n_T \int dt \int d\Omega \int dE \;\Phi_{\nu_{\tau}}(E,\theta) \; \sigma^{CC}_{\nu_{\tau}}(E) 
\end{eqnarray}
where $n_T$ is the number of target nucleons in the effective volume, and $\sigma^{CC}_{\nu_{\tau}}$ is the Charged Current cross-section, adequate in order to consider $\nu_{\tau}$ double-bang events. The function $\Phi_{\nu_{\tau}}(E,\theta)$ is the $\tau$-neutrino flux in the vicinity of the detector. We consider the number of events in the region $0^\circ <\theta<60^\circ$ around the nadir direction, for an observation time of ten years. We have taken the energy interval binning as $\Delta log_{10}E=0.25$.
To appreciate the size of the effect of Majorana neutrino production, we consider the percentage deviation between the non-standard and the $SM$ event numbers ($\Delta_{\%}=100\times (N_{\rm SM}-N_{\rm Maj})/N_{\rm SM}$), with $N_{\rm Maj}$ the number of events including the Majorana neutrino effects, and we compare it with the percentage relative error ($\delta_{\%}=100/\sqrt{N_{\rm SM}}$) for Poisson distributed events. The results are shown in Fig.\ref{fig:delta} for different values of the dominant coupling $\zeta_{4-f}$. The solid circles indicate the center of each energy bin. We consider the variations in this effective coupling due to the dominant contribution of the 4-fermion interactions to the deviation in the $\Phi_{\nu_{\tau}}$ flux.  As we can see from this figure, there is a region in the parameters space where the effect of Majorana neutrinos would be distinguishable from the $SM$ background, i.e. the percentage deviation $\Delta_{\%}$ is bigger than the SM error $\delta_{\%}$. 

In Fig.6 we show the region in the ($E_{\nu},\zeta_{4-f}$) plane where the studied phenomena could have a detectable impact. The region is limited by the curve for which $\Delta_{\%}$ equals the SM error $\delta_{\%}$, and the horizontal straightline, representing the upper bound for the 4-fermion coupling. As the Majorana effects decrease with lower energy, higher values for the $\zeta_{4-f}$ coupling are allowed. On the other hand, due to the spectral index ($-2.6$), the incident flux strongly decreases with growing energy, and this reduces the number of events, thus increasing the SM error. This gives bigger values for the effective coupling at higher energies, in order to have $\Delta_{\%}=\delta_{\%}$.

\begin{figure}
\begin{center}
\includegraphics[width=0.6\textwidth]{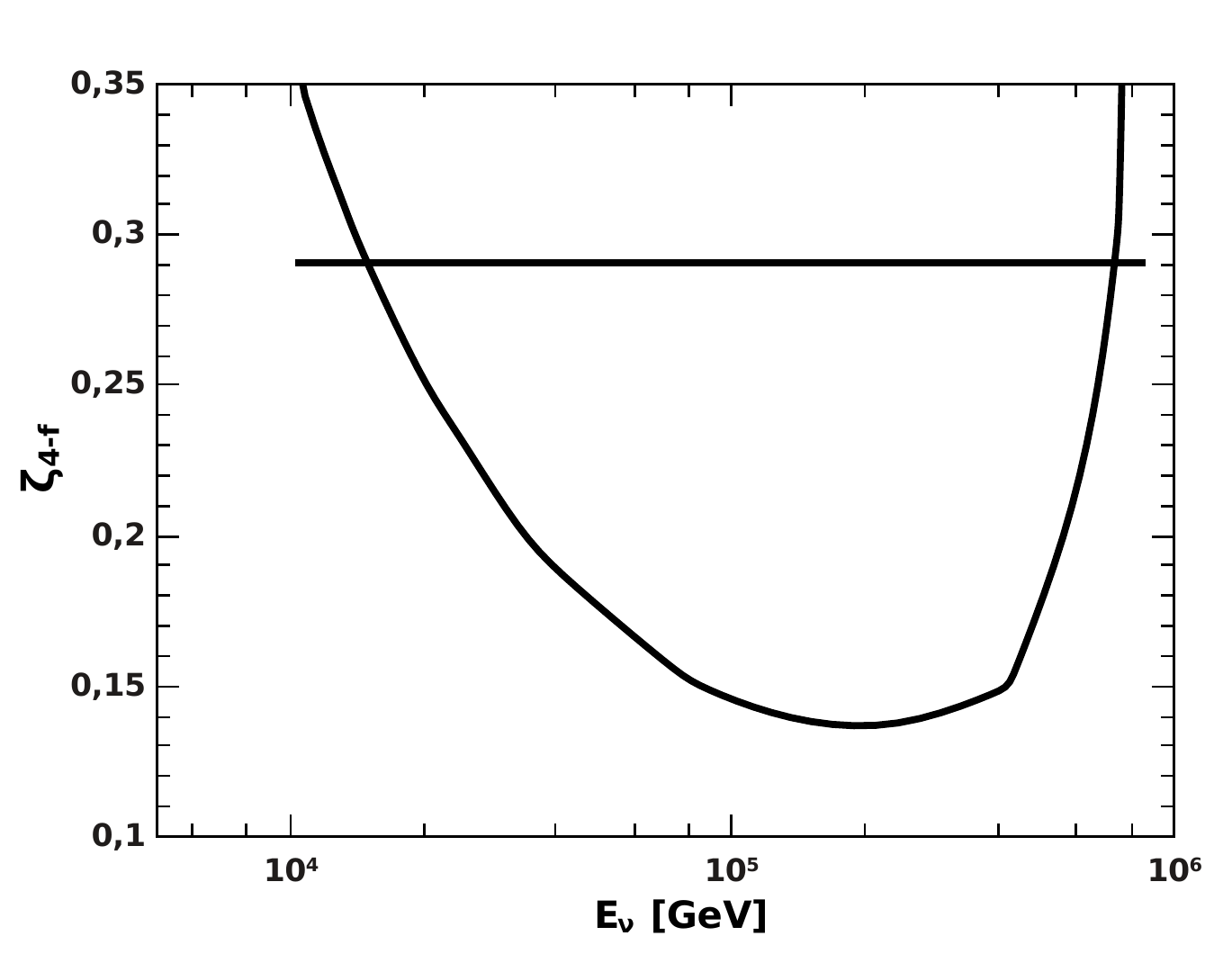}
%\hspace{4cm}
 \caption{\label{fig:reg_delta} Region in the ($E_{\nu},\zeta_{4-f}$) plane where the studied effects could have impact. The region is 
 limited for the curve where the percentage deviation $\Delta_{\%}$ equals the SM error $\delta_{\%}$ and a horizontal line which is the upper bound for the 4-fermion coupling
 Table \ref{tab:bounds}.} 
\end{center}
\end{figure}

\section{Final remarks \label{conclusiones}}
We have studied how the production of sterile Majorana neutrinos would affect the attenuation of cosmic $\nu_{\tau}$ neutrinos when
they pass through the Earth. For the propagation, we considered a system of transport equations for ordinary and Majorana neutrinos and the $\tau$ charged lepton, presenting our results for the flux attenuation with and without Majorana effects, and we show the percentage deviation between the $SM$ flux and the flux with Majorana attenuation. Our results can serve as a complementary tool to explore the effects of sterile neutrino physics, by directly studying  the effects of UHE neutrino interactions with the nucleons of the Earth using neutrino telescopes. Over the coming years, new neutrino telescopes are planned
to be working in the Northern hemisphere. In particular the European project KM3NeT \cite{Spurio:2012mh,DiCapuafortheKM3NeTCollaboration:2016ush,Piattelli:2015pmp}, originated in the projects ANTARES, NEMO and NESTOR will be installed in the Mediterranean sea with an instrumented volume of several cubic kilometers. This telescope along with the Baikal-GVD upgrade \cite{Avrorin:2012mg, Shaybonov:2015hcd} will improve the statistics, increasing the significance of the observations to bound new physics effects as the ones we discussed in this work. \\

{\bf Acknowledgements} 

We thank CONICET and Universidad Nacional de Mar del
Plata (Argentina); and PEDECIBA, ANII, and CSIC-UdelaR (Uruguay) for their 
financial supports.

\appendix

\section{N decay in the Laboratory}\label{ap1}
Here in this appendix we follow the development shown in the book of T.K.Gaisser \cite{Gaisser:1990vg}, in our case for the $N\rightarrow \gamma \nu$ decay.
First we obtain the $N$ decay width in its rest frame, and then boost the result to the Laboratory frame. In the $N$ rest frame we have the following expression: 
\begin{eqnarray}
\frac{1}{\Gamma_{\rm rest}}\frac{d\Gamma_{\rm rest}}{dx \, d\cos\theta_{\nu}}=  2\left(f_0(x)- P f_1(x)\cos\theta_{\nu}
\right),
\end{eqnarray}
where $\theta_\nu$ is the direction of motion of the final $\nu$ taken from the Majorana neutrino $N$ moving direction, and
$P=\cos\theta_P$ where $\theta_P$ is the angle between the Majorana neutrino spin direction in its rest frame, and its moving direction seen from the laboratory frame. The variable $x$ represents the quotient between the final neutrino energy in the rest frame of the $N$ and the mass of the Majorana
neutrino: $x=k^0/m_N$.
The functions $f_0(x)$ and $f_1(x)$ are
\begin{eqnarray}
f_0(x)= x (1-x)\delta(x-1/2) \nonumber \\ \nonumber
\\
f_1(x)=x^2 \delta(x-1/2),
\end{eqnarray}
To obtain the corresponding expression in the laboratory frame, we make the appropriate Lorentz transformations. 
Denoting by $E_\nu$ and $E_N$ the Laboratory energies of the final neutrino and the Majorana neutrino, respectively, we have
\begin{equation}
z=x(1-\beta_{N} \cos\theta_\nu),
\end{equation}
with $z=E_\nu/E_N$ and $\beta_{N}=\sqrt{1-{m_N}^2/E^2_{N}}\simeq 1$.

We implement the Lorentz transformation with the help of the $\delta$-function, yielding
\begin{eqnarray}
\frac{1}{\Gamma_{\rm LAB}} \frac{d\Gamma_{\rm LAB}}{dz \, dx  \, d\cos\theta_\nu} =  2 \left(f_0(x)-P f_1(x)
\cos\theta_\nu \right) \delta\left[z-x\left(1+\beta_N \cos\theta_{\nu}\right)\right].
\end{eqnarray}
We first integrate over $\theta_\nu$ and next we integrate over $x$ in the interval $(x_{\rm min},x_{\rm max})$ with $x_{\rm
min}=z/(1+\beta_N)$ and $x_{\rm max}={\rm min}(1,z/(1-\beta_N))$, obtaining
\begin{eqnarray}
\frac{1}{\Gamma_{\rm LAB}}\frac{d\Gamma_{\rm LAB}}{dz}=2(1-z)\Theta(1/2-x(z)_{min})\Theta(x(z)_{max}-1/2).
\end{eqnarray}

For the low mass range considered in this work the clearly dominant decay channel is the neutrino plus photon mode, and $\Gamma^{\rm
tot}_{LAB}(E)=\sum_{i=e,\mu,\tau}\Gamma^{N\rightarrow \nu_{i} \gamma}_{LAB}(E)$. Then we consider the $\nu_{\tau}$ decay channel, 
leading to the final $\nu_{\tau}$ neutrinos distribution in the laboratory frame:
\begin{eqnarray}
\frac{1}{\Gamma^{\rm tot}_{\rm LAB}(E)}\frac{d\Gamma^{N\rightarrow \nu_{\tau}\gamma}_{\rm LAB}}{dz} \equiv\frac{dn(z)}{dz}.
\end{eqnarray}
Thus, after the indicated integrations in the evolution equations, the useful expression that we obtain is
\begin{eqnarray}
\frac{dn(z)}{dz}=\frac{n(1-y)}{dy}= \frac{2}{3} y,
\end{eqnarray}
where $z=1-y$, $x_0=1/2$ and $P=+1$ for the right-handed Majorana neutrinos.
%\nmewpage
\bibliography{Bib_N_tierra_08_2016}

\end{document}